\documentclass[12pt]{article}
\usepackage{amsmath}
\usepackage{epsfig}

\def\no{\nonumber}
\def\eps{\varepsilon}
\def\Lib{\text{Li}_2}
\def\Lic{\text{Li}_3}

\def\Sab{\text{S}_{1,2}}

\def\Im{\text{Im}}
\def\as{\alpha_s}
\def\LQCD{\Lambda_\text{QCD}}
\def\MSbar{\ensuremath{\overline{\text{MS}}}}
\def\calS{\mathcal{S}}
\def\calP{\mathcal{P}}
\def\calO{\mathcal{O}}
\def\calC{\mathcal{C}}
\def\ubar{\bar{u}}
\def\SCETI{\ensuremath{\text{SCET}_\text{I}}}
\def\gev{\text{GeV}}
\def\slash#1{#1 \hskip-0.45em /}

\allowdisplaybreaks[3]

\begin{document}


\begin{titlepage}

\begin{flushright}
 LMU-ASC 32/07\\
 TTP07-08\\
 SFB/CPP-07-20
\end{flushright}
\vskip 2.2cm

\begin{center}
\Large\bf\boldmath
NNLO Vertex Corrections in charmless\\
hadronic $B$ decays: Imaginary part \unboldmath

\normalsize
\vskip 1.5cm

{\sc Guido~Bell\footnote{E-mail:bell@particle.uni-karlsruhe.de}}

\vskip .5cm

{\it Arnold Sommerfeld Center for Theoretical Physics, \\ 
Department f\"ur Physik, Ludwig-Maximilians-Universit\"at M\"unchen, \\ 
Theresienstra{\ss}e 37, D-80333 M\"unchen, Germany} 

\vskip .5cm

{\it Institut f\"ur Theoretische Teilchenphysik, \\
Universit\"at Karlsruhe, D-76128 Karlsruhe, Germany}  

\vskip 1.5cm

\end{center}

\begin{abstract}
\noindent
We compute the imaginary part of the 2-loop vertex corrections in the
QCD Factorization framework for hadronic two-body decays as
$B\to\pi\pi$. This completes the NNLO calculation of the imaginary part
of the topological tree amplitudes and represents an important step
towards a NNLO prediction of direct CP asymmetries in QCD
Factorization. Concerning the technical aspects, we find that soft and
collinear infrared divergences cancel in the hard-scattering kernels
which demonstrates factorization at the 2-loop order. All results are
obtained analytically including the dependence on the charm quark
mass. The numerical impact of the NNLO corrections is found to be
significant, in particular they lead to an enhancement of the strong
phase of the colour-suppressed tree amplitude.   
\end{abstract}

\vfill

\end{titlepage}


\section{Introduction}


Charmless hadronic $B$ decays provide important information on the
unitarity triangle which may help to reveal the nature of flavour mixing
and CP violation. In order to exploit the rich amount of data that is
currently being collected at the $B$ factories, a quantitative control
of the underlying strong-interaction effects is highly desirable. In the
QCD Factorization framework~\cite{BBNS} the hadronic matrix elements of
the operators in the effective weak Hamiltonian simplify considerably in
the heavy-quark limit. Schematically,        
\begin{align} 
\langle M_1 M_2 | Q_i | \bar{B} \rangle \; 
& \simeq \; F_+^{B M_1}(0) \; f_{M_2} \int du \; \; T_{i}^I(u) \;
  \phi_{M_2}(u) \no \\  
& \quad \; + \hat{f}_{B} \, f_{M_1} \, f_{M_2} \int d\omega dv du \; \;
  T_{i}^{II}(\omega,v,u) \; \phi_B(\omega) \; \phi_{M_1}(v) \;
  \phi_{M_2}(u), 
\label{eq:QCDF}
\end{align}
where the non-perturbative strong-interaction effects are encoded in a
form factor $F_+^{B M_1}$ at $q^2=0$, decay constants $f_M$ and
light-cone distribution amplitudes $\phi_M$. The short-distance kernels
$T_i^{I}=\calO(1)$ and $T_i^{II}=\calO(\as)$ provide the basis for a
systematic implementation of radiative corrections; the former contain
the short-distance interactions that do not involve the spectator
anti\-quark from the decaying $\bar{B}$ meson (\emph{vertex
  corrections}) and the latter describe the ones with the spectator
antiquark (\emph{spectator scattering}).   

The next-to-leading order (NLO) corrections to the kernels $T_i^{I,II}$,
which constitute an $\calO(\as)$ correction to naive factorization, are
already known from~\cite{BBNS}. Recently, the next-to-next-to-leading
order (NNLO) corrections to $T_i^{II}$ have been
computed~\cite{SpecScat:NLO:BJ05,SpecScat:NLO,SpecScat:NLO:BJ06,Jet,Jet:BY05}.
Due to the inter\-action with the soft spectator antiquark, the
spectator scattering term receives contributions from the hard scale
$\sim m_b$ and from an intermediate (hard-collinear) scale $\sim (\LQCD
m_b)^{1/2}$. Both types of contributions are now available at
$\calO(\as^2)$ (1-loop), indicating that the NNLO corrections are
numerically important.                 

In this work we compute NNLO corrections to $T_i^{I}$ for the so-called
topological tree amplitudes (which arise from the insertion of
current-current operators). In contrast to the spectator scattering
term, the vertex corrections are dominated solely by hard effects and
amount to a 2-loop calculation. In particular, we address the imaginary
part of the hard-scattering kernels which is the origin of a strong
rescattering phase shift that blurs the information on the weak
phases. As an imaginary part is first generated at $\calO(\as)$, higher
order perturbative corrections are expected to significantly influence
the pattern of strong phases and hence direct CP asymmetries. Our
calculation represents an important step towards a NNLO prediction of
direct CP asymmetries in QCD Factorization.           

The outline of this paper is as follows: In Section~\ref{sec:opbasis} we
present our strategy for the calculation of the topological tree
amplitudes by introducing two different operator
bases. Section~\ref{sec:2loop} contains the technical aspects of the
2-loop calculation. In Section~\ref{sec:subs} we show how to extract the
hard-scattering kernels from the hadronic matrix elements. Our
analytical results can be found in Section~\ref{sec:results}. The
numerical impact of the NNLO vertex corrections is investigated in
Section~\ref{sec:numerics} and we finally conclude in
Section~\ref{sec:conclusion}. A more detailed presentation of the
considered calculation can be found in \cite{GB:thesis}.

\newpage

\section{Choice of operator basis}

\label{sec:opbasis}

In view of the calculation of topological tree amplitudes, we restrict
our attention to the current-current operators of the effective weak
Hamiltonian for $b \to u$ transitions 
\begin{align}
\mathcal{H}_\text{eff} =
    \frac{G_F}{\sqrt{2}} \; V_{ub}V_{ud}^* \;
    \left( C_1 Q_1 + C_2 Q_2 \right)
    + \text{h.c.}
\label{eq:Heff}
\end{align}
Due to the fact that we work within Dimensional Regularization (DR), we
also have to consider evanescent operators~\cite{Evanescent}. These
non-physical operators vanish in four dimensions but contribute at
intermediate steps of the calculation in $d=4-2\eps$ dimensions. As the
imaginary part has effectively 1-loop complexity with respect to
renormalization at $\calO(\as^2)$, the considered calculation only
requires 1-loop evanescent operators. For our purposes the complete
operator basis is thus given by  
\begin{align}
\tilde Q_1 &=  
   \left[\bar u_i \gamma^\mu L\, b_i\right] \;
   \left[\bar d_j \gamma_\mu L\, u_j\right],\no\\
\tilde Q_2 &=  
   \left[\bar u_i \gamma^\mu L\, b_j\right] \;
   \left[\bar d_j \gamma_\mu L\, u_i\right],\no\\
\tilde E_1 &=  
   \left[\bar u_i \gamma^\mu\gamma^\nu\gamma^\rho L\,b_i\right] \;
   \left[\bar d_j \gamma_\mu\gamma_\nu\gamma_\rho L\, u_j\right]
   -(16-4\eps) \,\tilde Q_1,\no\\
\tilde E_2 &=  
   \left[\bar u_i \gamma^\mu\gamma^\nu\gamma^\rho L\,b_j\right] \;
   \left[\bar d_j \gamma_\mu\gamma_\nu\gamma_\rho L\, u_i\right]
   -(16-4\eps) \,\tilde Q_2,
\label{eq:TradBasis}
\end{align}
where $i,j$ are colour indices and $L=1-\gamma_5$. The operator basis in
(\ref{eq:TradBasis}) has been used in all previous calculations within
QCD
Factorization~\cite{BBNS,SpecScat:NLO:BJ05,SpecScat:NLO,SpecScat:NLO:BJ06}.
We refer to this basis as the \emph{traditional basis} for convenience
and denote the corresponding Wilson coefficients and operators with a
tilde.  

It has been argued by Chetyrkin, Misiak and M\"unz (CMM) that one should
use a different operator basis in order to perform multi-loop
calculations~\cite{CMM}. Although the deeper reason is related to the
penguin operators which we do not consider here, we prefer to introduce
the \emph{CMM basis} in view of future extensions of our work. This
basis allows to consistently use DR with a naive anticommuting
$\gamma_5$ to all orders in perturbation theory. In the CMM basis the
current-current operators and corresponding 1-loop evanescent operators
read (indicated by a hat) 
\begin{align}
\hat Q_1 &=  
  \left[\bar u_i \gamma^\mu L\, T^A_{ij} b_j\right] \;
  \left[\bar d_k \gamma_\mu L\, T^A_{kl} u_l\right],\no\\
\hat Q_2 &=  
  \left[\bar u_i \gamma^\mu L\, b_i\right] \;
  \left[\bar d_j \gamma_\mu L\, u_j\right],\no\\
\hat E_1 &=  
  \left[\bar u_i \gamma^\mu\gamma^\nu\gamma^\rho L\, T^A_{ij} b_j
  \right] \;
  \left[\bar d_k \gamma_\mu\gamma_\nu\gamma_\rho L\, T^A_{kl} u_l
  \right] 
  - 16 \,\hat Q_1,\no\\
\hat E_2 &=  
  \left[\bar u_i \gamma^\mu\gamma^\nu\gamma^\rho L\, b_i \right] \;
  \left[\bar d_j \gamma_\mu\gamma_\nu\gamma_\rho L\, u_j \right] 
  -16 \,\hat Q_2,
\label{eq:CMMBasis}
\end{align} 
with colour matrices $T^A$ and colour indices $i,j,k,l$.

Comparing the operator bases in (\ref{eq:TradBasis}) and
(\ref{eq:CMMBasis}) we observe two differences: First, the two bases use
different colour decompositions which is a rather trivial point. More
importantly, they contain slightly different definitions of evanescent
operators. Whereas the definitions in the CMM basis correspond to the
simplest prescription to define evanescent operators, subleading terms
of $\calO(\eps)$ appear in the one of the traditional basis. These terms
have been properly adjusted such that Fierz symmetry holds to 1-loop
order in $d$ dimensions.       

\begin{figure}[t!]
\centerline{\parbox{14cm}{\centerline{
\includegraphics[height=2.4cm]{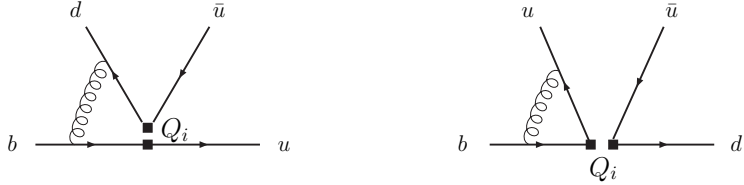}}
\caption{\label{fig:Insertions} \small \textit{Generic 1-loop diagram
    with different insertions of a four-quark operator $Q_i$. The upper
    lines go into the emitted meson $M_2$, the quark to the right of the
    vertex and the spectator antiquark in the $\bar B$ meson (not drawn)
    form the recoil meson $M_1$.}}}}   
\end{figure} 

We follow the notation of~\cite{BenekeNeubert} and express the hadronic
matrix elements of the effective weak Hamiltonian in terms of
topological amplitudes $\alpha_i(M_1 M_2)$. E.g.~the $B^-\to\pi^-\pi^0$
decay amplitude is written as 
\begin{align}
\sqrt{2} \; \langle \pi^- \pi^0 | \, \mathcal{H}_\text{eff} \, | B^-
\rangle &=  V_{ub}V_{ud}^* \; \big[\alpha_1(\pi\pi) + \alpha_2(\pi\pi)
\big] \; A_{\pi\pi}, 
\label{eq:amplitudes}
\end{align}  
where $A_{\pi\pi}=i G_F/\sqrt{2} \; m_B^2 F_+^{B\pi}(0)f_\pi$. The
amplitude $\alpha_{1}(M_1 M_2)$ is called the colour-allowed tree
amplitude which corresponds to the flavour content $[\bar{q}_s b]$ of
the decaying $\bar B$ meson, $[\bar{q}_s u]$ of the recoil meson $M_1$
and $[\bar{u} d]$ of the emitted meson $M_2$. The colour-suppressed tree
amplitude $\alpha_{2}(M_1 M_2)$ then belongs to the flavour contents
$[\bar{q}_s b]$, $[\bar{q}_s d]$ and $[\bar{u} u]$, respectively. For
more details concerning the definition of the topological amplitudes we
refer to section~2.2 in~\cite{BenekeNeubert}.   

According to this definition, the left (right) diagram in
Figure~\ref{fig:Insertions} contributes to the tree amplitude $\alpha_1$
($\alpha_2$). On the technical level these two insertions of a
four-quark operator correspond to two different calculations. Instead of
performing both calculations explicitly, we may alternatively compute
the amplitude $\alpha_2$ by inserting Fierz reordered operators into the
left diagram of Figure~\ref{fig:Insertions}. To do so, it is essential
to work with an operator basis that respects Fierz symmetry in $d$
dimensions. As we have argued above, Fierz symmetry indeed holds to
1-loop order in the traditional basis from (\ref{eq:TradBasis}) which
allows us to \emph{derive} $\alpha_2$ directly from $\alpha_1$. 

We conclude that the CMM basis is the appropriate choice for a 2-loop
calculation whereas the traditional basis provides a short-cut for the
derivation of the colour-suppressed amplitude. We therefore pursue the
following strategy for the calculation of the imaginary part of the NNLO
vertex corrections: We perform the explicit 2-loop calculation in the
CMM basis using the first type of operator insertion in the left diagram
from Figure~\ref{fig:Insertions}. From this we obtain $\alpha_1(\hat
C_i)$. We then transform this expression into the traditional basis
which yields $\alpha_1(\tilde C_i)$ and finally apply Fierz symmetry
arguments to derive $\alpha_2(\tilde C_i)$ from $\alpha_1(\tilde C_i)$
by simply exchanging $\tilde C_1\leftrightarrow \tilde C_2$.

\newpage

\section{2-loop calculation}

\label{sec:2loop}

\begin{figure}[p!]
\vspace{-2mm}
\centerline{\parbox{16cm}{
\includegraphics[width=16cm]{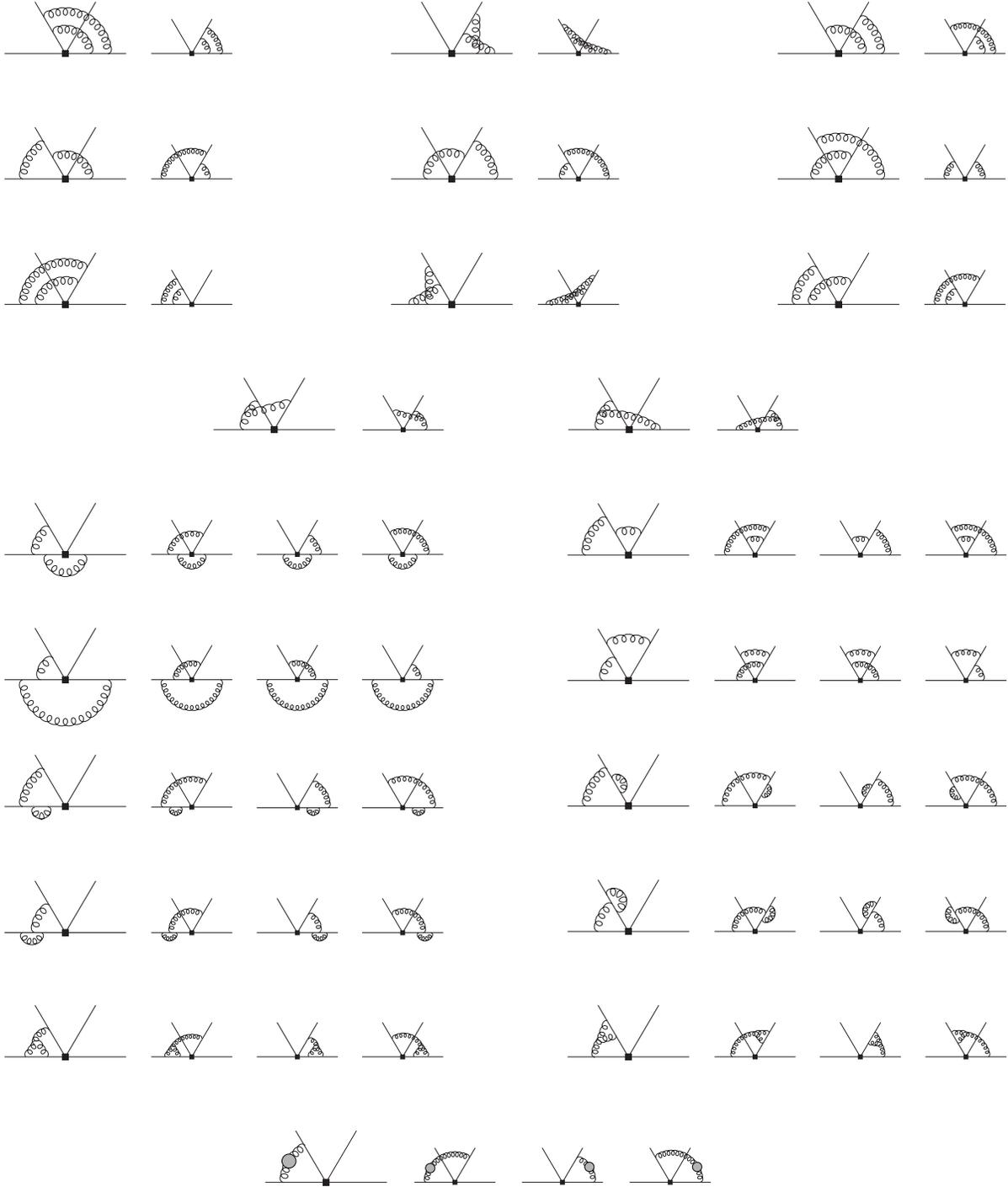}
\vspace{3mm}
\caption{\label{fig:Diagrams}  \small \textit{Full set of (naively)
    non-factorizable 2-loop diagrams. The bubble in the last four
    diagrams represents the 1-loop gluon self-energy. Only diagrams with
    final  state inter\-actions, i.e.~with at least one gluon connecting
    the line  to the right of the vertex with one of the upper lines,
    give rise to  an imaginary part.}}}}  
\end{figure}

The core of the considered calculation consists in the computation of
the matrix elements    
\begin{align}
\langle \hat{Q}_{1,2} \rangle \equiv 
\langle u(p') d(q_1) \bar u(q_2) | \hat{Q}_{1,2} | b(p) \rangle
\label{eq:partonicME}
\end{align}
to $\calO(\as^2)$ which represents a 2-loop calculation. As will be
described in Section~\ref{sec:FactNNLO}, only (naively) non-factorizable
diagrams with at least one gluon connecting the two currents in the left
diagram of Figure~\ref{fig:Insertions} have to be considered here. The
full NNLO calculation thus involves the 2-loop diagrams depicted in
Figure~\ref{fig:Diagrams}, but only about half of the diagrams give rise
to an imaginary part. It is an easy task to identify this subset of
diagrams since the generation of an imaginary part is always related to
final state interactions.     

In our calculation we treat the partons on-shell and write $q_1=u q$,
$q_2=\ubar q$ and $p'=p-q$ satisfying $q^2=0$ and $p^2=2p\cdot q=m_b^2$
(with $\ubar\equiv 1-u$). We use DR for the regularization of
ultraviolet (UV) and infrared (IR) divergences and an anticommuting
$\gamma_5$ according to the NDR scheme. We stress that we do not perform
any projection onto the bound states in the partonic calculation. We
instead treat the two currents in the four-quark operators independently
and make use of the equations of motion in order to simplify the Dirac
structures of the diagrams. In order to calculate the large number of
2-loop integrals we proceed as follows: Using a general tensor
decomposition of the loop integrals, we essentially deal with the
calculation of scalar integrals. With the help of an automatized
reduction algorithm, we are able to express several thousands of scalar
integrals in terms of a small set of so-called Master Integrals
(MIs). The most difficult part finally consists in the calculation of
these MIs.    

In the remainder of this section we present some techniques that we have
found useful for the considered calculation. We sketch the basic ideas
of the aforementioned reduction algorithm and discuss several techniques
for the calculation of the MIs. We refer to the references quoted in the
following subsections for more detailed descriptions (see
also~\cite{GB:thesis}).  

\subsection{Reduction to Master Integrals}

Any scalar 2-loop integral in our calculation can be expressed as
\begin{align}
I(u) \;=\;
        \int d^d k \, d^d l \quad
        \frac{\calS_1^{m_1} \ldots \calS_s^{m_s}}
        {\calP_1^{n_1}\ldots \calP_p^{n_p}},
\label{eq:Int}
\end{align}
where the $\calS_i$ are scalar products of a loop momentum with an
external momentum or of two loop momenta. The $\calP_i$ denote the
denominators of propagators and the exponents fulfil $n_i, m_i \geq
0$. The scalar integrals themselves depend on the convolution variable
$u$ in the factorization formula (\ref{eq:QCDF}). Very few integrals,
arising from diagrams with a charm quark in a closed fermion loop,
depend in addition on the ratio $z\equiv m_c/m_b$. We have suppressed
this dependence in (\ref{eq:Int}) for simplicity.   

Notice that an integral has different representations in terms of
$\{\calS,\calP,n,m\}$ because of the freedom to shift loop momenta in
DR. It is the underlying topology, i.e. the inter\-connection of
propagators and external momenta, which uniquely defines the
integral. In the following we loosely use the word topology in order
to classify the integrals. An integral with $t$ different propagators
$\calP_i$ with $n_i>0$ is called a $t$-topology. The integrals in the
considered calculation have topology $t\leq6$.  

The reduction algorithm makes use of various identities which relate
integrals with different exponents $\{n,m\}$. The most important class
of identities are the integration-by-parts identities~\cite{IBP} which
follow from the fact that surface terms vanish in DR  
\begin{align}
\int d^d k \, d^d l \quad
        \frac{\partial}{\partial v^\mu} \; \;
        \frac{\calS_1^{m_1} \ldots \calS_s^{m_s}}
        {\calP_1^{n_1}\ldots \calP_p^{n_p}}  \;=\; 0,
        \qquad \qquad v\in\{k,l\}.
\label{eq:IBP}
\end{align}
In order to obtain scalar identities we may contract (\ref{eq:IBP}) with
any loop or external momentum under the integral before performing the
derivative. In our case of two loop and two external momenta we generate
in this way eight identities from each integral.  

A second class of identities, called Lorentz-invariance
identities~\cite{LI}, exploits the fact that the integrals in
(\ref{eq:Int}) transform as scalars under a Lorentz-transformation of
the external momenta. In this way we may generate up to six identities
from each integral depending on the number of external momenta. In our
example with only two linearly independent external momenta $p$ and $q$
there is only one such identity given by  
\begin{align}
\int d^d k \, d^d l \;
        \left[ p \cdot q \left(p^\mu \frac{\partial}{\partial p^\mu} -
            q^\mu \frac{\partial}{\partial q^\mu} \right) + q^2 \, p^\mu
          \frac{\partial}{\partial q^\mu} - p^2 \, q^\mu
          \frac{\partial}{\partial p^\mu} \right] \;
        \frac{\calS_1^{m_1} \ldots \calS_s^{m_s}} 
        {\calP_1^{n_1}\ldots \calP_p^{n_p}}  \;=\; 0. 
\label{eq:LI}
\end{align}
In total we obtain nine identities from a given integral, each of the
identities containing the integral itself, simpler integrals with
smaller $\{n,m\}$ and more complicated integrals with larger
$\{n,m\}$. It is important that the number of identities grows faster
than the number of unknown integrals for increasing $\{n,m\}$. Hence,
for large enough $\{n,m\}$ the system of equations becomes (apparently)
overconstrained and can be used to express more complicated integrals in
terms of simpler ones. Not all of the identities being linearly
independent, some integrals turn out to be irreducible to which we refer
as MIs.    

In the considered calculation we typically deal with systems of
equations made of several thousands equations. The solution being
straight-forward, the runtime of the reduction algorithm depends
strongly on the order in which the equations are solved. As a guideline
for an efficient implementation we have followed the algorithm
from~\cite{Laporta}.   
\begin{figure}[t!]
\centerline{\parbox{14cm}{\centerline{
\includegraphics[width=14cm]{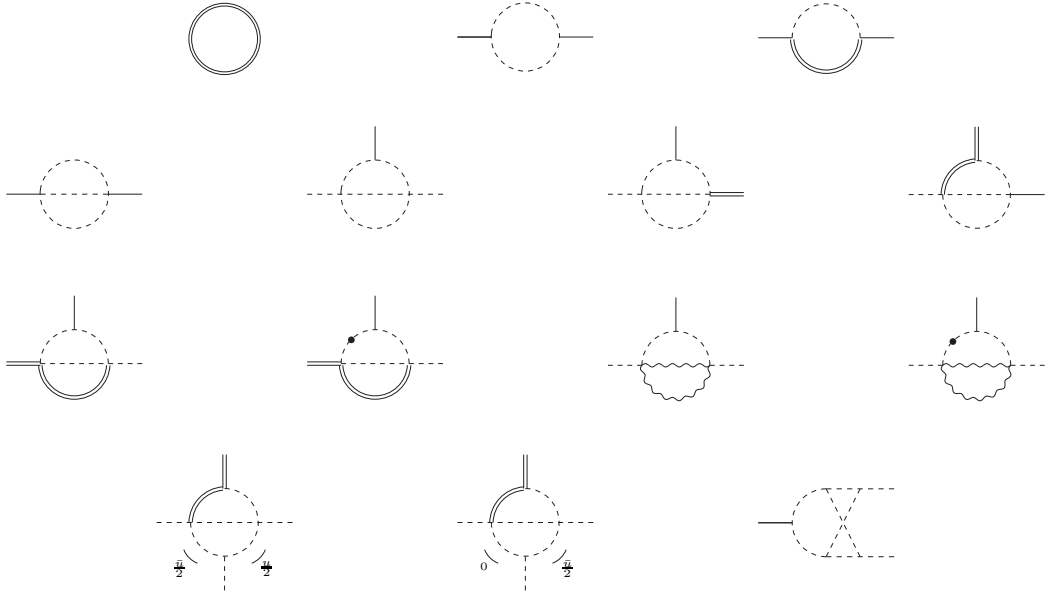}}
\caption{\label{fig:MIs} \small \textit{Scalar Master Integrals that
    appear in our calculation. We use dashed lines for massless
    propagators and double (wavy) lines for the ones with mass $m_b$
    ($m_c$). Dashed/solid/double external lines correspond to
    virtualities $0\,/\,u m_b^2\,/\,m_b^2$, respectively. Dotted
    propagators are taken to be squared.}}}} 
\end{figure}

The reduction algorithm enables us to express the diagrams of
Figure~\ref{fig:Diagrams} as linear combinations of MIs which are
multiplied by some Dirac structures. As the coefficients in these linear
combinations are real, we may extract the imaginary part of a diagram at
the level of the MIs which is a much simpler task than for the full
diagrams. As depicted in Figure~\ref{fig:MIs}, we find 14 MIs that
contribute to the calculation of the imaginary part of the NNLO vertex
corrections. 

\subsection{Calculation of Master Integrals}

Some MIs in Figure~\ref{fig:MIs} can be solved easily e.g.~with the help
of Feynman parameters. For the more complicated MIs the method of
differential equations~\cite{DiffEqs} in combination with the formalism
of Harmonic Polylogarithms (HPLs)~\cite{HPLs} turned out to be very
useful. In this section we give brief reviews of these techniques and
conclude with a comment on the calculation of the boundary conditions to
the differential equations. 

The analytical results for the MIs from Figure~\ref{fig:MIs} can be
found in Appendix~A.1 of~\cite{GB:thesis}. As an independent check of
our results we evaluated the MIs numerically using the method of sector
decomposition~\cite{SecDecomp}.  

\subsubsection*{Method of Differential Equations}

The MIs are functions of the physical scales of the process which are
given by scalar products of the external momenta and masses of the
particles. In our calculation the MIs depend on the dimensionless
variable $u$ as in (\ref{eq:Int}). 

For a given MI we perform the derivative with respect to $u$ and
interchange the order of integration and derivation 
\begin{align}
\frac{\partial}{\partial u} \; \text{MI}_{\,i}(u)
        \;=\;
        \int d^d k \, d^d l \quad
        \frac{\partial}{\partial u} \; \;
        \frac{\calS_1^{m_1} \ldots \calS_s^{m_s}}
        {\calP_1^{n_1}\ldots \calP_p^{n_p}}.
\end{align}
The right-hand side being of the same type as equations (\ref{eq:IBP})
and (\ref{eq:LI}), this procedure again leads to a sum of various
integrals with different exponents $\{n,m\}$. With the help of the
reduction algorithm, these integrals can be expressed in terms of MIs
which yields a differential equation of the form   
\begin{align}
\frac{\partial}{\partial u} \; \text{MI}_{\,i}(u)
        \;=\;
        a(u;d) \; \text{MI}_{\,i}(u)
        + \sum_{j \neq i} \,b_j(u;d) \; \text{MI}_{\,j}(u),
\label{eq:DiffEq}
\end{align}
where we indicated that the coefficients $a$ and $b_j$ may depend on the
dimension $d$. The inhomogeneity of the differential equation typically
contains MIs of subtopologies which are supposed to be known in a
bottom-up approach. In case of the MIs from the third line of
Figure~\ref{fig:MIs}, one MI in the inhomogeneous part is of the same
topology as the MI on the left hand side of (\ref{eq:DiffEq}) and thus
unknown. Writing down the differential equation for this MI, we find
that we are left with a coupled system of linear, first order
differential equations.   

We are looking for a solution of the differential equation in terms of
an expansion   
\begin{align}
\text{MI}_{\,i}(u) \;=\; \sum_j \, \frac{c_{ij}(u)}{\eps^j}.
\label{eq:ansatz}
\end{align}
Expanding (\ref{eq:DiffEq}) then gives much simpler differential
equations for the coefficients $c_{ij}$ which can be solved order by
order in $\eps$. In addition, in the case where we were left with a
coupled system of differential equations, the system turns out to
decouple in the expansion. The solution of the homogeneous equations is
in general straight-forward. The inhomogeneous equations can then be
addressed with the  method of the variation of the constant. This in
turn leads to indefinite integrals over the inhomogeneities which
typically contain products of rational functions with logarithms or
related functions as dilogarithms. With the help of the formalism of
HPLs these integrations simplify substantially.  

\subsubsection*{Harmonic Polylogarithms}

The formalism of HPLs~\cite{HPLs} allows to rewrite the integrations
mentioned at the end of the last section in terms of familiar
transcendental functions which are defined by repeated integration over
a set of basic functions. We briefly summarize their basic features
here, focussing on the properties that are relevant for our calculation.

The HPLs, denoted by $H(\vec{m}_w;x)$, are described by a
$w$-dimensional vector $\vec{m}_w$ of parameters and by its argument
$x$. We restrict our attention to the parameters $0$ and $1$ in the
following. The basic definitions of the HPLs are for weight $w=1$  
\begin{align}
H(0;x) \equiv \ln x, \qquad \qquad
H(1;x) \equiv - \ln (1-x)
\end{align}
and for weight $w>1$
\begin{align}
H(a, \vec{m}_{w\text{--}1};x) &\equiv
        \int_0^x dx' \; f(a;x') \; H(\vec{m}_{w\text{--}1};x'),
\label{eq:HPL}
\end{align}
where the basic functions $f(a;x)$ are given by
\begin{align}
f(0;x) = \frac{\partial}{\partial x} H(0;x) = \frac{1}{x}, \qquad \qquad
f(1;x) = \frac{\partial}{\partial x} H(1;x) = \frac{1}{1-x}.
\end{align}
In the case of $\vec{m}_w=\vec{0}$, the definition in (\ref{eq:HPL})
does not apply and the HPLs read  
\begin{align}
H(0, \ldots,0;x) &\equiv \frac{1}{w!} \ln^w x.
\end{align}
The HPLs form a closed and linearly independent set under integrations
over the basic functions $f(a;x)$ and fulfil an algebra such that a
product of two HPLs of weight $w_1$ and $w_2$ gives a linear combination
of HPLs of weight $w=w_1+w_2$. 

As described above, the solution of the differential equations leads to
integrals over products of some rational functions with some
transcendental functions as logarithms or dilogarithms. More precisely,
we find e.g.~integrals of the type   
\begin{align}
\int^x dx' \;
\left\{ \frac{1}{1-x'},\, \frac{1}{x'^2},\, 1 \right\}  \;
H(\vec{m}_{w};x'). 
\label{eq:HPLints}
\end{align}
It turns out that all these integrals can be expressed as linear
combinations of HPLs of weight $w+1$. This is obvious for the first
integral as it simply corresponds to the definition of a HPL,
cf.~(\ref{eq:HPL}) with $a=1$. For the other integrals in
(\ref{eq:HPLints}), an integration-by-parts leads either to a recurrence
relation or again to integrals of the type (\ref{eq:HPL}). Not all
integrals in our calculation fall into the simple pattern
(\ref{eq:HPLints}), but a large part of this calculation can be
performed along these lines.      

In the considered calculation we encounter HPLs of weight $w\leq 3$. Our
results can be expressed in terms of the following minimal set of HPLs
\begin{align}
H(0;x) &= \ln x,  &
H(0,0,1;x) &= \Lic(x), 
\no \\
H(1;x) &= - \ln (1-x), &
H(0,1,1;x) &= \Sab(x).
\no \\
H(0,1;x) &= \Lib(x),
\end{align}
The situation is more complicated for the last two MIs in the third line
of Figure~\ref{fig:MIs} where the internal charm quark introduces a new
scale. However, a closer look reveals that these MIs depend on two
physical scales only, namely $u m_b^2$ and $m_c^2 = z^2 m_b^2$. The MIs
can then be solved within the formalism of HPLs in terms of the ratio
$\xi \equiv z^2/u$ if we allow for more complicated arguments of the
HPLs as e.g.~$\eta \equiv\frac12 \left( 1- \sqrt{1+4\xi} \right)$.   

\subsubsection*{Boundary conditions}

A unique solution of a differential equation requires the knowledge of
its boundary conditions. In the considered calculation  the boundary
conditions typically represent single-scale integrals corresponding to
$u=0$ or $1$. It is of crucial importance that the integral has a smooth
limit at the chosen point such that setting $u=0$ or $u=1$ does not
modify the divergence structure introduced in (\ref{eq:ansatz}).   

In some cases the methods described so far can also be applied for the
calculation of the boundary conditions since setting $u=0$ or $1$ leads
to simpler topologies which may turn out to be reducible. If so, the
reduction algorithm can be used to express the integral in terms of
known MIs. If not, a different strategy is mandatory. In this case we
tried to calculate the integral with the help of Feynman parameters and
managed in some cases to express the integral in terms of hypergeometric
functions which we could expand in $\eps$ with the help of the {\sc
  Mathematica} package {\sc HypExp}~\cite{HypExp}. Finally, the most
difficult single-scale integrals could be calculated with Mellin-Barnes
techniques~\cite{MellinBarnes}.


\section{Renormalization and IR subtractions}

\label{sec:subs}

The matrix elements $\langle \hat{Q}_{i} \rangle$ which we obtained from
computing the 2-loop diagrams in Figure~\ref{fig:Diagrams} are
ultraviolet (UV) and infrared (IR) divergent. In this section we show
how to extract the finite hard-scattering kernels $T_i^I$ from these
matrix elements.        

\subsection{Renormalization}

The renormalization procedure involves standard QCD counterterms, which
amount to the calculation of various 1-loop diagrams, as well as
counterterms from the effective Hamiltonian. We write the renormalized
matrix elements as  
\begin{align}
\langle \hat{Q}_{i} \rangle_\text{ren} &= Z_\psi \, \hat{Z}_{i j} \, 
\langle \hat{Q}_{j} \rangle_\text{bare}, 
\label{eq:Qren}
\end{align}
where $Z_\psi \equiv Z_b^{1/2} Z_q^{3/2}$ contains the wave-function
renormalization factors $Z_b$ of the $b$-quark and $Z_q$ of the massless
quarks, whereas $\hat{Z}$ is the operator renormalization matrix in the
effective theory. We introduce the following notation for the
perturbative expansions of these quantities 
\begin{align}
\langle \hat{Q}_{i} \rangle_\text{ren/bare} =
    \sum_{k=0}^\infty \left( \frac{\as}{4\pi} \right)^k \langle
    \hat{Q}_{i} \rangle_\text{ren/bare}^{(k)}, \hspace{2.2cm} \no \\ 
Z_\psi = 1 + \sum_{k=1}^\infty  \left( \frac{\as}{4\pi} \right)^k
Z_\psi^{(k)}, \hspace{1.5cm} \hat{Z}_{i j} = \delta_{ij} +
\sum_{k=1}^\infty  \left( \frac{\as}{4\pi} \right)^k
\hat{Z}_{ij}^{(k)}
\end{align}
and rewrite (\ref{eq:Qren}) in perturbation theory up to NNLO which
yields 
\begin{align}
\langle \hat{Q}_{i} \rangle_\text{ren}^{(0)} &=
    \langle \hat{Q}_{i} \rangle_\text{bare}^{(0)}, \no \\
\langle \hat{Q}_{i} \rangle_\text{ren}^{(1)} &=
    \langle \hat{Q}_{i} \rangle_\text{bare}^{(1)}
    + \left[ \hat{Z}_{ij}^{(1)} + Z_\psi^{(1)} \delta_{ij} \right]
    \langle \hat{Q}_{j} \rangle_\text{bare}^{(0)}, \no \\
\langle \hat{Q}_{i} \rangle_\text{ren}^{(2)} &=
    \langle \hat{Q}_{i} \rangle_\text{bare}^{(2)}
    + \left[ \hat{Z}_{ij}^{(1)} + Z_\psi^{(1)} \delta_{ij} \right] \langle \hat{Q}_{j} \rangle_\text{bare}^{(1)}
    + \left[ \hat{Z}_{ij}^{(2)} + Z_\psi^{(1)} \hat{Z}_{ij}^{(1)} +
      Z_\psi^{(2)} \delta_{ij} \right] \langle \hat{Q}_{j}
    \rangle_\text{bare}^{(0)}. 
\label{eq:expQren} 
\end{align}
The full calculation thus requires the operator renormalization matrices
$\hat{Z}^{(1,2)}$. For the calculation of the imaginary part, the terms
proportional to the tree level matrix elements do not contribute and
$\hat{Z}^{(2)}$ drops out in (\ref{eq:expQren}) as expected for an
effective 1-loop calculation.    

Mass and wave function renormalization are found to be higher order
effects. For the renormalization of the coupling constant we use 
\begin{align}
Z_g = 1 - \frac{\as}{4\pi \eps} \left( \frac{11}{2} - \frac13 n_f
\right) + \calO(\as^2). 
\end{align}
according to the \MSbar-scheme. The expression for the 1-loop
renormalization matrix $\hat{Z}^{(1)}$ can be found e.g.~in 
\cite{Gambino:2003zm} and reads 
\renewcommand{\arraycolsep}{2mm}
\begin{align}
\hat{Z}^{(1)} &= \left(
\begin{array}{c c c c}
\rule[-2mm]{0mm}{7mm} -2 & \frac43 & \frac{5}{12} & \frac{2}{9} \\
\rule[-2mm]{0mm}{7mm} 6 & 0 & 1 & 0
\end{array}
\right) \, \frac{1}{\eps}, 
\end{align}
where the two lines correspond to the basis of physical operators $\{
\hat{Q}_1, \hat{Q}_2\}$ and the four columns to the extended basis $\{
\hat{Q}_1, \hat{Q}_2, \hat{E}_1, \hat{E}_2 \}$ including the evanescent
operators $\hat{E}_{1,2}$ defined in (\ref{eq:CMMBasis}). 

\subsection{Factorization in NNLO}
\label{sec:FactNNLO}

In this section it will be convenient to introduce the following
short-hand notation for the factorization formula (\ref{eq:QCDF})  
\begin{align}
\langle \hat{Q}_{i} \rangle_\text{ren} &= F \cdot T_i \otimes \Phi +
\ldots 
\label{eq:FFshort}
\end{align}
where $F$ denotes the $B\to M_1$ form factor, $T_i$ the hard-scattering
kernels $T^I_i$ and $\Phi$ the product of the decay constant $f_{M_2}$
and the distribution amplitude $\phi_{M_2}$. The convolution in
(\ref{eq:QCDF}) has been represented by the symbol $\otimes$ and the
ellipsis contain the terms from spectator scattering which we disregard
in the following.  

\begin{figure}[t!]
\centerline{\parbox{14cm}{\centerline{
\includegraphics[width=12cm]{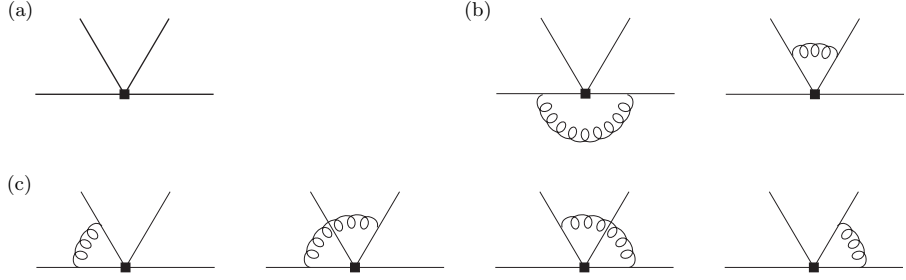}} 
\caption{\label{fig:LONLO} \small \textit{Tree level diagram~(a),
    naively factorizable~(b) and non-factorizable~(c) NLO diagrams.}}}} 
\end{figure}

Formally, we may introduce the perturbative expansions
\begin{align}
F = \sum_{k=0}^\infty \left( \frac{\as}{4\pi} \right)^k F^{(k)},
\hspace{1cm} 
T_i = \sum_{k=0}^\infty \left( \frac{\as}{4\pi} \right)^k T_i^{(k)},
\hspace{1cm} 
\Phi = \sum_{k=0}^\infty \left( \frac{\as}{4\pi} \right)^k \Phi^{(k)}. 
\end{align}
Up to NNLO the expansion of (\ref{eq:FFshort}) then yields
\begin{align}
\label{eq:expFFshort}
\langle \hat{Q}_{i} \rangle_\text{ren}^{(0)} &=
    F^{(0)} \cdot T_i^{(0)} \otimes \Phi^{(0)}, \no \\ 
\langle \hat{Q}_{i} \rangle_\text{ren}^{(1)} &= 
    F^{(0)} \cdot T_i^{(1)} \otimes \Phi^{(0)} + F^{(1)} \cdot T_i^{(0)}
    \otimes \Phi^{(0)} + F^{(0)} \cdot T_i^{(0)} \otimes \Phi^{(1)}, \no
    \\ 
\langle \hat{Q}_{i} \rangle_\text{ren}^{(2)} &=
    F^{(0)} \cdot T_i^{(2)} \otimes \Phi^{(0)} + F^{(1)} \cdot T_i^{(1)}
    \otimes \Phi^{(0)} + F^{(0)} \cdot T_i^{(1)} \otimes \Phi^{(1)} \no
    \\ 
    & \quad + F^{(2)} \cdot T_i^{(0)} \otimes \Phi^{(0)} + F^{(1)} \cdot
    T_i^{(0)} \otimes \Phi^{(1)} + F^{(0)} \cdot T_i^{(0)} \otimes
    \Phi^{(2)}. 
\end{align}
In LO the comparison of (\ref{eq:expQren}) and (\ref{eq:expFFshort})
gives the trivial relation   
\begin{align}
    \langle \hat{Q}_{i} \rangle^{(0)} \equiv
    \langle \hat{Q}_{i} \rangle_\text{bare}^{(0)} =
    F^{(0)} \cdot T_i^{(0)} \otimes \Phi^{(0)}
\end{align}
which states that the LO kernels $T_i^{(0)}$ can be computed from the
tree level diagram in Figure~\ref{fig:LONLO}a. In order to address
higher order terms we split the matrix elements into its (naively)
factorizable (f) and non-factorizable (nf) contributions   
\begin{align}
\langle \hat{Q}_{i} \rangle_\text{bare}^{(k)} &\equiv
    \langle \hat{Q}_{i} \rangle_\text{f}^{(k)} +
    \langle \hat{Q}_{i} \rangle_\text{nf}^{(k)}.
\end{align}
The corresponding 1-loop diagrams are shown in Figure~\ref{fig:LONLO}b
and \ref{fig:LONLO}c respectively. To this order (\ref{eq:expQren}) and
(\ref{eq:expFFshort}) lead to 
\begin{align}
    &\langle \hat{Q}_{i} \rangle_\text{f}^{(1)} + \langle \hat{Q}_{i}
    \rangle_\text{nf}^{(1)} 
    + \left[ \hat{Z}_{ij}^{(1)} + Z_\psi^{(1)} \delta_{ij} \right]
    \langle \hat{Q}_{j} \rangle^{(0)} \no \\ 
    & \qquad = F^{(0)} \cdot T_i^{(1)} \otimes \Phi^{(0)} + F^{(1)}
    \cdot T_i^{(0)} \otimes \Phi^{(0)} + F^{(0)} \cdot T_i^{(0)} \otimes
    \Phi^{(1)}, 
\end{align}
which splits into
\begin{align}
    \langle \hat{Q}_{i} \rangle_\text{nf}^{(1)}
    + \hat{Z}_{ij}^{(1)} \langle \hat{Q}_{j} \rangle^{(0)}
    &= F^{(0)} \cdot T_i^{(1)} \otimes \Phi^{(0)}
\label{eq:NLOFact}
\end{align}
for the calculation of the NLO kernels $T_i^{(1)}$ and
\begin{align}
    \langle \hat{Q}_{i} \rangle_\text{f}^{(1)} + Z_\psi^{(1)} \langle
    \hat{Q}_{i} \rangle^{(0)} 
    &= F^{(1)} \cdot T_i^{(0)} \otimes \Phi^{(0)} + F^{(0)} \cdot
    T_i^{(0)} \otimes \Phi^{(1)}, 
\end{align}
which shows that the factorizable diagrams and the wave-function
renormalization are absorbed by the form factor and wave function
corrections $F^{(1)}$ and $\Phi^{(1)}$.  

This suggests in NNLO the following structure
\begin{align}
    & \langle \hat{Q}_{i} \rangle_\text{f}^{(2)} + Z_\psi^{(1)} \langle
    \hat{Q}_{i} \rangle_\text{f}^{(1)} + Z_\psi^{(2)} \langle
    \hat{Q}_{i} \rangle^{(0)} \no \\ 
    & \qquad = F^{(2)} \cdot T_i^{(0)} \otimes \Phi^{(0)} + F^{(1)}
    \cdot T_i^{(0)} \otimes \Phi^{(1)} + F^{(0)} \cdot T_i^{(0)} \otimes
    \Phi^{(2)}. 
\end{align}
These terms are thus irrelevant for the calculation of the NNLO kernels
$T_i^{(2)}$ which justifies that we could restrict our attention to the
non-factorizable 2-loop diagrams from Figure~\ref{fig:Diagrams}. In NNLO
the remaining terms from (\ref{eq:expQren}) and (\ref{eq:expFFshort})
contain non-trivial IR subtractions    
\begin{align}
    & \langle \hat{Q}_{i} \rangle_\text{nf}^{(2)} + Z_\psi^{(1)} \langle
    \hat{Q}_{i} \rangle_\text{nf}^{(1)} + \hat{Z}_{ij}^{(1)} \left[
      \langle \hat{Q}_{j} \rangle_\text{nf}^{(1)} + \langle \hat{Q}_{j}
      \rangle_\text{f}^{(1)} \right] + \left[ \hat{Z}_{ij}^{(2)} +
      Z_\psi^{(1)} \hat{Z}_{ij}^{(1)} \right] \langle \hat{Q}_{j}
    \rangle^{(0)}  \no \\ 
& \qquad = F^{(0)} \cdot T_i^{(2)} \otimes \Phi^{(0)} + F^{(1)} \cdot
T_i^{(1)} \otimes \Phi^{(0)} + F^{(0)} \cdot T_i^{(1)} \otimes
\Phi^{(1)}. 
\end{align}
This equation can be simplified further when we make the wave function
renormalization factors in the form factor and the distribution
amplitude explicit   
\begin{align}
    F = Z_b^{1/2} Z_q^{1/2} F_\text{amp}, \hspace{2cm}
    \Phi = Z_q \, \Phi_\text{amp}.
\label{eq:amp}
\end{align}
Notice that the resulting amputated form factor $F_\text{amp}$ and wave
function $\Phi_\text{amp}$ contain UV-divergences by construction. Using
(\ref{eq:NLOFact}), we see that the wave function renormalization
factors cancel and arrive at the final formula for the calculation of
the NNLO kernels $T_i^{(2)}$ 
\begin{align}
    & \langle \hat{Q}_{i} \rangle_\text{nf}^{(2)}
    + \hat{Z}_{ij}^{(1)} \left[ \langle \hat{Q}_{j}
      \rangle_\text{nf}^{(1)} + \langle \hat{Q}_{j}
      \rangle_\text{f}^{(1)} \right] 
    + \hat{Z}_{ij}^{(2)}  \langle \hat{Q}_{j} \rangle^{(0)}  \no \\ 
    & \qquad = F^{(0)} \cdot T_i^{(2)} \otimes \Phi^{(0)}
        + F_\text{amp}^{(1)} \cdot T_i^{(1)} \otimes \Phi^{(0)} 
        + F^{(0)} \cdot T_i^{(1)} \otimes \Phi_\text{amp}^{(1)}. 
\label{eq:Master}
\end{align}
As the tree level matrix elements and the factorizable 1-loop diagrams
do not give rise to an imaginary part, these terms can be disregarded in
the present calculation. 

\subsection{IR subtractions}

We now consider the IR subtractions on the right hand side of
(\ref{eq:Master}) in some detail. Let us first address the NLO kernels
$T_i^{(1)}$ which can be determined from equation
(\ref{eq:NLOFact}). The renormalization in the evanescent sector implies
that the left hand side of (\ref{eq:NLOFact}) is free of contributions
from evanescent operators \emph{up to the finite order}
$\eps^0$. However, as the NLO kernels enter (\ref{eq:Master}) in 
combination with the form factor correction $F_\text{amp}^{(1)}$ which
contains double (soft and collinear) IR divergences, the NLO kernels are
required here up to $\calO(\eps^2)$. Concerning the subleading terms of
$\calO(\eps)$, the evanescent operators do not drop out on the left hand
side of (\ref{eq:NLOFact}) and we therefore have to extend the
factorization  formula on the right hand side to include these
evanescent structures as well. Schematically,   
\begin{align}
    \langle \hat{Q}_{i} \rangle_\text{nf}^{(1)}
    + \hat{Z}_{ij}^{(1)} \langle \hat{Q}_{j} \rangle^{(0)}
    &= F^{(0)} \cdot T_i^{(1)} \otimes \Phi^{(0)} + 
    F_E^{(0)} \cdot T_{i,E}^{(1)} \otimes \Phi_E^{(0)}
\label{eq:NLOFact:Evan}
\end{align}
with a kernel $T_{i,E}^{(1)}=\calO(1)$ and an evanescent tree level
matrix element $F_E^{(0)} \Phi_E^{(0)}=\calO(\eps)$\footnote{In the
  notation of~\cite{SpecScat:NLO:BJ05}, the right hand side of
  (\ref{eq:NLOFact:Evan}) corresponds to matrix elements of
  \SCETI~operators of the form $[(\bar{\xi} W_{c1}) \Gamma_1 h_v]
  [(\bar{\chi} W_{c2}) \Gamma_2 (W^\dagger_{c2} \chi)]$. In NNLO we
  match onto two \SCETI~operators with Dirac-structures $\Gamma_1
  \otimes \Gamma_2$ given by $O_1= \slash{n}_+ L \otimes
  \frac{\slash{n}_-}{2} L$ and $O_2= \slash{n}_+ \gamma_\perp^\mu
  \gamma_\perp^\nu L \otimes \frac{\slash{n}_-}{2} {\gamma_\perp}_\mu
  {\gamma_\perp}_\nu  L$ (in our notation $p'=\frac12m_b n_-$ and
  $q=\frac12 m_b n_+$). The matrix element of
  $O_1$  
  defines our structure $F^{(0)} \Phi^{(0)}$ and the evanescent
  combination $3O_2-12O_1$ defines $F_E^{(0)}
  \Phi_E^{(0)}$.}. Similarly, the right hand side of (\ref{eq:Master})
has to be modified to include these evanescent structures    
\begin{align}
    & \langle \hat{Q}_{i} \rangle_\text{nf}^{(2)}
    + \hat{Z}_{ij}^{(1)} \left[ \langle \hat{Q}_{j}
      \rangle_\text{nf}^{(1)} + \langle \hat{Q}_{j}
      \rangle_\text{f}^{(1)} \right] 
    + \hat{Z}_{ij}^{(2)}  \langle \hat{Q}_{j} \rangle^{(0)}  \no \\ 
    & \qquad = F^{(0)} \cdot T_i^{(2)} \otimes \Phi^{(0)}
        + F_\text{amp}^{(1)} \cdot T_i^{(1)} \otimes \Phi^{(0)} 
        + F^{(0)} \cdot T_i^{(1)} \otimes \Phi_\text{amp}^{(1)} \no \\
    & \qquad + F_E^{(0)} \cdot T_{i,E}^{(2)} \otimes \Phi_E^{(0)}
        + F_\text{amp,E}^{(1)} \cdot T_{i,E}^{(1)} \otimes \Phi_E^{(0)} 
        + F_E^{(0)} \cdot T_{i,E}^{(1)} \otimes \Phi_\text{amp,E}^{(1)}.
\label{eq:MasterEvan}
\end{align}     
Notice that the term with the kernel $T_{i,E}^{(2)}=\calO(1)$ is not
required to extract the finite piece of the physical NNLO kernel
$T_i^{(2)}$.   

From the calculation of the 1-loop diagrams in Figure~\ref{fig:LONLO}c,
we find that the NLO kernels vanish in the colour-singlet case,
$T_2^{(1)}=T_{2,E}^{(1)}=0$, whereas the imaginary part of the
colour-octet kernels is given by    
\begin{align} 
\frac{1}{\pi} \; \Im \;T_1^{(1)}(u)  &= \frac{C_F}{2N_c}
\bigg\{ (-3-2\ln u+2\ln \ubar) \Big( 1+\eps L +\frac12 \eps^2 L^2\Big)
\no \\ 
& \hspace{1.6cm} 
- ( 11 - 3 \ln \ubar  - \ln^2 u+ \ln^2 \ubar ) \Big( \eps +\eps^2 L
\Big) \no \\  
& \hspace{1.6cm} 
+ \bigg[ \frac{3\pi^2}{4} - 26 + \Big( 2 + \frac{\pi^2}{2} \Big) \ln u +
\Big( 9 - \frac{\pi^2}{2} \Big) \ln \ubar \no \\ 
& \hspace{2.2cm} - 
\frac32 \ln^2 \ubar - \frac13 \left( \ln^3 u - \ln^3 \ubar \right)
\bigg] \eps^2 + \calO(\eps^3) \bigg\}, \nonumber  
\end{align}
\begin{align} 
\frac{1}{\pi} \; \Im \;T_{1,E}^{(1)}(u)  &= - \frac{C_F}{4N_c} \bigg\{
1+\eps L + \Big( \frac83 - \frac12 \ln u -\frac 12 \ln \ubar \Big) \eps
+  \calO(\eps^2) \bigg\},  
\label{eq:T1}
\end{align}
where $L \equiv \ln \mu^2/m_b^2$ and we recall that $\ubar\equiv1-u$. 

\subsubsection*{Form factor subtractions}

\begin{figure}[b!]
\centerline{\parbox{13cm}{
\centerline{\includegraphics[width=4.2cm]{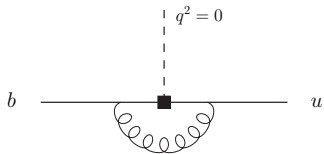}}
\caption{\label{fig:FFcorr}  \small \textit{1-loop contribution to the
    form factor correction $F^{(1)}_\text{amp}$.}}}} 
\end{figure}

We now address the form factor corrections which require the calculation
of the diagram in Figure~\ref{fig:FFcorr} (for on-shell quarks) and its
counterterm. According to the definition of $F_\text{amp}$ in
(\ref{eq:amp}), we do not have to consider the wave function
renormalization of the quark fields.      

We again have to compute the corrections for physical and evanescent
operators. Concerning the physical operator, the counterterm is found to
vanish which reflects the conservation of the vector current. The
evaluation of the 1-loop diagram in Figure~\ref{fig:FFcorr} gives  
\begin{align}
F_\text{amp}^{(1)} \,\Phi^{(0)}= - C_F \left(
  \frac{e^{\gamma_E}\mu^2}{m_b^2} \right)^\eps \Gamma(\eps) \;
\frac{1-\eps+2\eps^2}{\eps(1-2\eps)}  \,\; F^{(0)}  \,\Phi^{(0)},
\label{eq:FFphys}
\end{align}
which contains double IR singularities as mentioned at the beginning of
this section. On the other hand, the 1-loop diagram with an insertion of
the evanescent operator yields a contribution proportional to the
evanescent \emph{and} the physical operators. We now have to adjust the
counterterm such that the renormalized (IR-finite) matrix element of the
evanescent operator vanishes (which ensures that the evanescent
structures disappear in the final factorization formula). We obtain  
\begin{align}
F_\text{amp,E}^{(1)} \,\Phi_E^{(0)}&= C_F \left[ \left(
    \frac{e^{\gamma_E}\mu^2}{m_b^2} \right)^\eps  \Gamma(\eps) \;
  \frac{24 \eps (1+\eps)}{(1-\eps)^2} - 24 \right] \;  F^{(0)}
\,\Phi^{(0)}  \nonumber\\
& \hspace{4mm} 
-  C_F \left( \frac{e^{\gamma_E}\mu^2}{m_b^2} \right)^\eps  \Gamma(\eps)
\; \frac{1-3\eps+\eps^2+3\eps^3+2\eps^4}{\eps(1-2\eps)(1-\eps)^2} \,\;
F_E^{(0)} \,\Phi_E^{(0)}.  
\label{eq:FFevan}
\end{align}
The form factor subtractions in (\ref{eq:MasterEvan}) then follow from
combining (\ref{eq:T1}), (\ref{eq:FFphys}) and (\ref{eq:FFevan}). We
emphasize that the corrections related to the evanescent operator
\emph{do not} induce a contribution to the physical NNLO kernel
$T_1^{(2)}$ in this case since   
\begin{align}
\frac{1}{\pi} \; F_\text{amp,E}^{(1)} \; \Im \;T_{1,E}^{(1)} \; 
\Phi_E^{(0)} \quad \to \quad
\bigg[\calO(\eps) \bigg] \;  F^{(0)} \,\Phi^{(0)}.
\end{align}   

\subsubsection*{Wave function subtractions}

\begin{figure}[t!]
\centerline{\parbox{13cm}{
\centerline{\includegraphics[width=12cm]{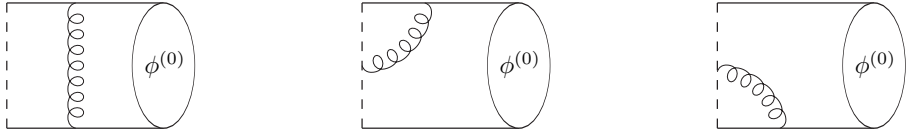}}
\caption{\label{fig:WFcorr}  \small \textit{1-loop contributions to the
    wave function correction $\Phi^{(1)}_\text{amp}$. The dashed line
    indicates the Wilson-line connecting the quark and antiquark
    fields.}}}}     
\end{figure}

Concerning the wave function corrections we are left with the
calculation of the diagrams in Figure~\ref{fig:WFcorr} for collinear and
on-shell partons with momenta $u q$ and $\ubar q$. However, as in our
set-up $q^2=0$ all these diagrams vanish due to scaleless integrals in
DR. We conclude that the wave function corrections are determined
entirely by the counter\-terms. We compute these counterterms by
calculating the diagrams from Figure~\ref{fig:WFcorr} with an
(IR-finite) off-shell regularization prescription in order to isolate
the UV-divergences (which are independent of the IR regulator). The
counter\-term for the physical operator is found to be       
\begin{align}
F^{(0)} \, \Phi_\text{amp}^{(1)} (u) = - \frac{2 C_F}{\eps}
\int_0^1 dw \; V(u,w) \; F^{(0)} \, \Phi^{(0)} (w)
\label{eq:Phiphys} 
\end{align}
with the familiar Efremov-Radyushkin-Brodsky-Lepage (ERBL)
kernel~\cite{ERBL} 
\begin{align}
V(u,w) = \left[ \theta(w-u) \frac{u}{w} \left(1 + \frac{1}{w-u}
\right) + \theta(u-w) \frac{\bar u}{\bar w} \left(1 + \frac{1}{\bar
w-\bar u} \right) \right]_+
\end{align}
where the plus-distribution is defined as
$\left[ f(u,w) \right]_+ =  f(u,w) - \delta(u-w) \int_0^1 dv \;
f(v,w)$.
For the evanescent operator we obtain
\begin{align}
F_E^{(0)} \, \Phi_\text{amp,E}^{(1)} (u) = - \frac{2 C_F}{\eps} \int_0^1
dw \, \bigg[ 24 \eps \,V_E(u,w) \; F^{(0)} \, \Phi^{(0)} (w) + V(u,w) \;
F_E^{(0)} \, \Phi_E^{(0)} (w) \bigg] 
\label{eq:Phievan} 
\end{align}
where $V_E(u,w)$ denotes the spin-dependent part of the ERBL kernel
given by 
\begin{align}
V_E(u,w) = \theta(w-u) \frac{u}{w}  + \theta(u-w) \frac{\bar u}{\bar w}.
\end{align}
Notice that the evanescent operators \emph{do} induce a finite
contribution to the physical NNLO kernel $T_1^{(2)}$ in this case as the
convolution with the corresponding NLO kernel implies  
\begin{align}
\frac{1}{\pi} \; F_E^{(0)} \; \Im \;T_{1,E}^{(1)} \; 
\Phi_\text{amp,E}^{(1)} \quad \to \quad 
\bigg[ \frac{6C_F^2}{N_c} + \calO(\eps) \bigg] \;  F^{(0)}
\,\Phi^{(0)}. 
\end{align} 
We finally quote the result for the convolution with the physical NLO
kernel 
\begin{align}
&\frac{1}{\pi} \; F^{(0)} \; \Im \;T_1^{(1)} \; \Phi_\text{amp}^{(1)}
\;=\; \frac{C_F^2}{N_c} \bigg\{ \bigg[ \frac{\pi^2}{3} + \frac{\ln
  u}{\ubar} - \frac{\ln \ubar}{u}  + \ln^2 u - 2 \ln u \ln \ubar - \ln^2
\ubar -4 \Lib (u) \bigg] \no \\ 
&\hspace{4.5cm} 
\bigg( \frac{1}{\eps} + L \bigg) + \frac{\pi^2}{2} - \frac{15}{4} - 2
\zeta_3 + \frac{5u-4}{2} \bigg( \frac{\ln u}{\ubar} + \frac{\ln
  \ubar}{u} \bigg)- \frac{\pi^2}{3} \ln \ubar    \no \\ 
& \hspace{4.5cm}  
- 3 \Lib(u) - \frac{1}{2 \ubar} \ln^2 u + \frac{1-3u}{2u}\ln^2 \ubar -
\frac23 \ln^3 u  + \ln^2 u \ln \ubar  \no \\ 
& \hspace{4.5cm} 
+ \frac23 \ln^3 \ubar+ 2 \ln \ubar \, \Lib (u) + 2 \Lic(u) + 2 \Sab (u)
+\calO(\eps) \bigg\} \, F^{(0)} \, \Phi^{(0)}. 
\end{align}


\section{Vertex Corrections in NNLO}

\label{sec:results}

We now have assembled all pieces required for the calculation of the
NNLO kernels $T_i^{(2)}$ from (\ref{eq:MasterEvan}). We indeed observe that
all UV and IR divergences cancel in the hard-scattering kernels as
predicted by the QCD Factorization framework. Since this is the result
of a complicated subtraction procedure, this can also be seen as a very
stringent cross-check of our calculation.      

\subsection{$\alpha_1$ in CMM basis}

The procedure outlined so far leads to the colour-allowed tree amplitude
in the CMM operator basis defined in (\ref{eq:CMMBasis}). We write  
\begin{align}
\alpha_1(M_1 M_2) &= \hat{C}_2 +\frac{\as}{4\pi} \, \frac{C_F}{2 
N_c} \bigg[ \hat{C}_1 \hat{V}^{(1)} + \frac{\as}{4\pi} \left( 
\hat{C}_{1} \, \hat{V}_1^{(2)} + \hat{C}_{2} \, 
\hat{V}_2^{(2)}\right) + \calO(\as^2)  \bigg] + \ldots 
\label{eq:alpha1Mod}
\end{align}
where the ellipsis denote the terms from spectator scattering which are
irrelevant for our purposes. In the CMM basis, the imaginary part of the
vertex corrections $\hat{V}^{(1,2)}$ takes the form     
\begin{align}
\frac{1}{\pi} \; \Im \;\hat{V}^{(1)}   &\equiv
    \int_0^1 du \; g_1(u)  \; \phi_{M_2}(u), \no\\
\frac{1}{\pi} \; \Im \;\hat{V}_1^{(2)} &\equiv
    \int_0^1 du \; \bigg\{ \Big[ \Big( \frac{29}{3} C_A - \frac23 n_f
    \Big) g_1(u) + C_F \,h_1(u) \Big] \ln \frac{\mu^2}{m_b^2}  \no \\ 
    & \hspace{3mm} + C_F \, h_2(u) + C_A \, h_3(u) + (n_f-2) \, h_4(u;0) +
    h_4(u;z) + h_4(u;1) \bigg\} \,  \phi_{M_2}(u), \no \\  
    \frac{1}{\pi} \; \Im \;\hat{V}_2^{(2)} &\equiv  \int_0^1 du \;
    \bigg\{ -6\, g_1(u)\, \ln \frac{\mu^2}{m_b^2} + h_0(u) \bigg\} \,
    \phi_{M_2}(u). 
\label{eq:V12mod}
\end{align}
In writing (\ref{eq:V12mod}) we have made the dependence on the
renormalization scale explicit and disentangled contributions that
belong to different colour structures. The NLO kernel $g_1(u)$ is given
by  
\begin{align}
g_1(u) &= -3 -2 \ln u + 2 \ln \ubar
\label{eq:kernels:g}
\end{align}
and the NNLO kernels $h_i(u)$ will be specified below. The kernel
$h_4(u;z_f)$ stems from diagrams with a closed fermion loop and depends
on the mass of the internal quark through $z_f=m_f/m_b$. We keep a
non-zero charm quark mass and write $z\equiv z_c = m_c/m_b$ for
simplicity.   

\newpage
The NNLO kernels were so far unknown. They are found in this work to be 
\begin{align}
h_0(u)  &=  
\bigg[ \frac{155}{4} + 4 \zeta_3 + 4 \Lic(u) - 4\Sab(u) -12 \ln u\,
\Lib(u) +\frac43 \ln^3 u -6\ln^2 u\ln \ubar \no \\  
&\hspace{8mm} + \frac{2-u^2}{\ubar}\Lib(u)
-\frac{5-3u+3u^2-2u^3}{2\ubar}\ln^2 u +\frac{3-2u^4}{2u\ubar}\ln u
\ln\ubar  \no \\  
&\hspace{8mm} - \Big( \frac{4-11u+2u^2}{\ubar}+\frac{4\pi^2}{3}\Big) \ln
u - \frac{(5+6u^2-12u^4)\pi^2}{24u \ubar} +
(u\leftrightarrow\ubar)\bigg]  \no \\  
&\hspace{5mm} + \bigg[ \frac{3-u+7u^2}{2\ubar}\ln^2 u -
\frac{11-10u^2}{4u\ubar} \Lib(u)+ \frac{1-14u^2}{4\ubar}\ln u \ln \ubar
\no \\  
&\hspace{12mm} + \frac{46-51u}{\ubar}\ln u -
\frac{(41-42u^2)\pi^2}{24\ubar} - (u\leftrightarrow\ubar) \bigg], \no \\
h_1(u) &=
36 + \bigg[ 2 \ln^2 u  - 4 \Lib(u) + \frac{2(13-12u)}{\ubar} \ln u
-(u\leftrightarrow\ubar) \bigg], \no\\  
h_2(u) &= 
\bigg[ 79 + 32 \zeta_3 -16 \Lic(u) - 32 \Sab(u) + \frac83 \ln^3 u
+\frac{2(4-u^2)}{\ubar} \Lib(u) \no \\ 
&\hspace{8mm} - \frac{13-9u+6u^2-4u^3}{2\ubar}\ln^2 u +
\frac{17-6u^2-8u^4}{4u\ubar}\ln u \ln \ubar \no \\ 
&\hspace{8mm} - 2\Big(\frac{5-11u+2u^2}{\ubar}+\frac{4\pi^2}{3} \Big)
\ln u- \frac{(1+14u^2-8u^4)\pi^2}{8u\ubar} +
(u\leftrightarrow\ubar)\bigg] \no \\ 
&\hspace{5mm} + \bigg[ 4 \Lic(u) + 4\Sab(u) -\frac23 \ln^3 u +2 \ln^2 u
\ln \ubar -\frac{9-14u^2}{u\ubar} \Lib(u) \no \\ 
&\hspace{12mm}+ \frac{13-11u+14u^2}{2\ubar} \ln^2
u+\frac{5-7u^2}{\ubar}\ln u \ln \ubar  \no \\ 
&\hspace{12mm}+ 4 \Big( \frac{24-23u}{\ubar} + \frac{\pi^2}{3} \Big) \ln
u - \frac{(26-21u^2)\pi^2}{6\ubar} - (u\leftrightarrow\ubar) \bigg],\no
\\ 
h_3(u) &= 
\bigg[ - \frac{1379}{24} - 12 \zeta_3 + 6 \Lic(u) +12 \Sab(u) -\ln^3 u -
\frac{4-u^2}{\ubar} \Lib(u) \no \\  
&\hspace{8mm} + \frac{9-2u+6u^2-4u^3}{4\ubar}\ln^2 u-
\frac{7+4u^2-4u^4}{4u\ubar}\ln u \ln \ubar \no \\
&\hspace{8mm} + \Big( \frac{41-66u + 8u^2}{4\ubar}+ \pi^2 \Big) \ln u +
\frac{(1+6u^2-4u^4)\pi^2}{8u\ubar} + (u\leftrightarrow\ubar)\bigg]\no\\
&\hspace{5mm} + \bigg[ -2 \Lic(u) + 4 \Sab(u) + 4 \ln u \, \Lib(u) +
\frac13 \ln^3 u + \frac{15-26u^2}{4u\ubar} \Lib(u) \no \\
&\hspace{12mm}+ \frac{11-14u-42u^2}{12\ubar} \ln^2 u -
\frac{11-14u^2}{4\ubar} \ln u \ln \ubar \no\\
&\hspace{12mm}- \Big( \frac{2165-2156u}{36\ubar} - \frac{\pi^2}{3} \Big)
\ln u + \frac{(53-42u^2)\pi^2}{24\ubar} - (u\leftrightarrow\ubar)
\bigg],\no\\ 
h_4(u;z) &= 
\bigg[ \frac{17}{6} + \frac{7\xi}{\ubar} -2 \xi^2 \ln^2 \frac{x_1}{x_2}
+ \frac{\xi}{\ubar} \ln z^2 -(1+2\xi)\ln u  \no \\ 
&\hspace{6mm} +\Big( 2(1+4\xi) x_1 + 4\xi x_2 \Big) \ln x_1 - \Big(
4\xi x_1+ 2 (1+4\xi) x_2 \Big) \ln x_2  + (u\leftrightarrow\ubar)\bigg]
\no \\ 
&\hspace{0mm} + \bigg[ \frac{94z^2}{9\ubar} - \frac{2(1-3\xi^2)}{3}
\ln^2 \frac{x_1}{x_2} -\frac43 \ln u \ln z^2 + \frac{(1-2u)(6\ubar-u
  \xi^2)}{9u\ubar\xi} \ln z^2  \no \\ 
&\hspace{6mm} +\frac{12+29\xi+2\xi^2}{9\xi} \ln u
- \frac{2}{9\xi} \Big( (1+4\xi)(6+5\xi) x_1 -6
(1-3\xi^2) x_2 \Big) \ln x_1 \no \\
&\hspace{6mm} - \frac{2}{9\xi} \Big( 6 (1-3\xi^2) x_1 - (1+4\xi)(6+5\xi)
x_2 \Big) \ln x_2 - (u\leftrightarrow\ubar) \bigg],
\label{eq:kernels:h}
\end{align}
where the last kernel has been given in terms of 
\begin{align}
x_{1} \equiv \frac12 \left( \sqrt{1+4\xi} - 1 \right), \qquad x_{2}
\equiv \frac12 \left( \sqrt{1+4\xi} + 1 \right), \qquad \xi \equiv
\frac{z^2}{u}. 
\end{align}
In the massless limit $h_4(u;z)$ becomes
\begin{align}
h_4(u;0) &=
    \frac{17}{3} - \frac23 \ln^2 u + \frac23 \ln^2 \ubar +\frac{20}{9} 
    \ln u - \frac{38}{9} \ln \ubar. 
\end{align}

\subsection{$\alpha_1$ and $\alpha_2$ in traditional basis}

Following our strategy from Section~\ref{sec:opbasis}, we compute the
colour-suppressed amplitude $\alpha_2$ by rewriting the colour-allowed
amplitude $\alpha_1$ in the traditional operator basis from
(\ref{eq:TradBasis}). Manifest Fierz symmetry in this basis relates the
two amplitudes via        
\begin{align}
\alpha_i(M_1 M_2) &= \tilde{C}_i + \frac{\tilde{C}_{i\pm1}}{N_c}  
+\frac{\as}{4\pi} \, \frac{C_F}{N_c} 
\bigg[ \tilde{C}_{i\pm1} \tilde{V}^{(1)} 
+ \frac{\as}{4\pi} \left( 
  \tilde{C}_{i} \, \tilde{V}_1^{(2)} 
+ \tilde{C}_{i\pm1} \, \tilde{V}_2^{(2)}
\right) + \calO(\as^2)  \bigg] + \ldots
\label{eq:alpha12Trad}
\end{align}
where the upper (lower) signs apply for $i=1$ ($i=2$) and the ellipsis
correspond to the terms from spectator scattering. In order to derive
$\tilde{V}^{(1,2)}$ we have to transform the Wilson coefficients in the
CMM basis $\hat{C}_{i}$ into the ones of the traditional basis
$\tilde{C}_{i}$. To NLL approximation this transformation can be found
e.g.~in~\cite{CMM} and reads      
\begin{align}
\hat{C}_1 &= 2 \tilde{C}_2 + \frac{\as}{4\pi} \left( 4 \tilde{C}_1 +
  \frac{14}{3} \tilde{C}_2 \right) + \calO(\as^2),\no \\ 
\hat{C}_2&= \tilde{C}_1 + \frac13 \tilde{C}_2 +
\frac{\as}{4\pi} \left( \frac{16}{9} \tilde{C}_2 \right) + \calO(\as^2).
\label{eq:WilsonTransform}
\end{align}
Combining (\ref{eq:alpha1Mod}), (\ref{eq:alpha12Trad}) and
(\ref{eq:WilsonTransform}) we obtain  
\begin{align}
\frac{1}{\pi} \; \Im \;\tilde{V}^{(1)}   &= 
    \frac{1}{\pi} \; \Im \;\hat{V}^{(1)} \no\\
  &= \int_0^1 du \; g_1(u)  \; \phi_{M_2}(u), \no\\
\frac{1}{\pi} \; \Im \;\tilde{V}_1^{(2)} &=
    \frac{1}{\pi} \; \Im \bigg[ \frac12 \,\hat{V}^{(2)}_2 +2
    \,\hat{V}^{(1)} \bigg] \no\\ 
  &= \int_0^1 du \; \bigg\{ -3\, g_1(u) \ln \frac{\mu^2}{m_b^2} +2
  \,g_1(u) +\frac12 \, h_0(u)  \bigg\} \,  \phi_{M_2}(u), \no \\ 
\frac{1}{\pi} \; \Im \;\tilde{V}_2^{(2)} &=
    \frac{1}{\pi} \; \Im \bigg[ \hat{V}^{(2)}_1 
    + \Big( \frac{C_A}{2} - C_F \Big) \hat{V}^{(2)}_2 
    + \Big( 4C_F - C_A \Big) \hat{V}^{(1)} \bigg] \no\\ 
  &= \int_0^1 du \; \bigg\{ \Big[ 
  \Big( \frac{20}{3} C_A + 6 C_F - \frac23 n_f \Big) g_1(u) +
  C_F h_1(u) \Big] \ln \frac{\mu^2}{m_b^2}  \no \\
  & \hspace{3mm}
  + C_F \bigg[ h_2(u) - h_0(u) +4g_1(u) \bigg]
  + C_A \bigg[ h_3(u) + \frac12 h_0(u) -g_1(u) \bigg] \no \\ 
  & \hspace{3mm} 
  + (n_f-2) \, h_4(u;0) + h_4(u;z) + h_4(u;1) \bigg\} \, \phi_{M_2}(u).
\label{eq:V12trad}
\end{align}
Equation (\ref{eq:V12trad}) represents the central result of our
analysis, specifying the imaginary part of the colour-allowed tree
amplitude $\alpha_1$ and the colour-suppressed tree amplitude $\alpha_2$
according to (\ref{eq:alpha12Trad}). The expression for
$\tilde{V}^{(1)}$ is in agreement with~\cite{BBNS}, whereas the
expressions for $\tilde{V}^{(2)}_{1,2}$ are new. The kernels $g_1(u)$
and $h_i(u)$ can be found in (\ref{eq:kernels:g}) and
(\ref{eq:kernels:h}), respectively. The terms proportional to $n_f$ have
already been considered in the analysis of the large $\beta_0$-limit in
\cite{betazero}. Our results are in agreement with these findings.

\subsection{Convolutions in Gegenbauer expansion}

Our results in (\ref{eq:V12trad}) have been given in terms of
convolutions of hard-scattering kernels with the light-cone distribution
amplitude of the emitted meson $M_2$. We may explicitly calculate these
convolution integrals by expanding the distribution amplitude into the
eigenfunctions of the 1-loop evolution kernel   
\begin{align}
\phi_{M_2}(u) &=
     6 u \ubar \left[ 1 + \sum_{n=1}^\infty \, a_n^{M_2} \;
     C_n^{(3/2)}(2u-1) \right], 
\label{eq:Gegenbauer}
\end{align}
where $a_n^{M_2}$ and $C_n^{(3/2)}$ are the Gegenbauer moments and
polynomials, respectively. We truncate the Gegenbauer expansion at $n=2$
and perform the convolution integrals analytically. We find  
\begin{align}
\int_0^1 du \; g_1(u)  \; \phi_{M_2}(u) &=
    -3 -3 \, a_1^{M_2}, \hspace{7cm}\no \\
\int_0^1 du \; h_0(u)  \; \phi_{M_2}(u) &=
    \frac{1333}{12} + \frac{47\pi^2}{45} -16 \zeta_3 +\left(
      \frac{15}{4} + \frac{23\pi^2}{5}  \right) a_1^{M_2}  \no\\ 
  & \hspace{6mm} - \left(  \frac{173}{30} + \frac{18\pi^2}{35}\right)
  a_2^{M_2} ,\no \\
\int_0^1 du \; h_1(u)  \; \phi_{M_2}(u) &= 
    36 + 28 \, a_1^{M_2},\no \\
\int_0^1 du \; h_2(u)  \; \phi_{M_2}(u) &=
    \frac{1369}{6} + \frac{139\pi^2}{45}  -32 \zeta_3 - \left(
      \frac{17}{6} - \frac{51\pi^2}{5}  \right) a_1^{M_2} \no \\ 
    & \hspace{6mm} - \left(\frac{103}{15} + \frac{71\pi^2}{35} \right)
    a_2^{M_2},\no \\ 
\int_0^1 du \; h_3(u)  \; \phi_{M_2}(u) &=
    - \frac{481}{3} + \frac{7\pi^2}{30} +12 \zeta_3 - \left(
      \frac{643}{12} +\frac{11\pi^2}{10} \right) a_1^{M_2} \no \\ 
    & \hspace{6mm} - \left( \frac{1531}{80} - \frac{169\pi^2}{70}
    \right) a_2^{M_2}, \no \\
H_4(z) \equiv \int_0^1 du \; h_4(u;z)  \; \phi_{M_2}(u) &\no\\
& \hspace{-4.5cm}
= \frac{22}{3} +148 z^2-96 z^4 F(z) -36 z^4 \ln^2 \frac{y_1}{y_2} \no \\
& \hspace{-3.3cm} 
-2\Big[1-(2y_1+1)(1+22z^2)\Big] \ln \frac{y_1}{y_2} -4\ln y_2 \no \\ 
& \hspace{-4cm}
 +\bigg\{ 7 + 164 z^2 + 180z^4 + 144 z^6 - 288z^4 F(z) +12z^4 (3 + 16 z^2
 + 12 z^4) \ln^2 \frac{y_1}{y_2} \no \\  
& \hspace{-3.3cm}    
-2\Big[1 - (2y_1 + 1)(1 + 22z^2 + 84z^4 + 72z^6)\Big] \ln
\frac{y_1}{y_2} -4 \ln y_2\bigg\} \, a_1^{M_2}\no \\   
& \hspace{-4cm}
+\bigg\{ \frac35 + 244 z^2 + \frac{148}{3} z^4 - 640 z^6 - 960 z^8  +24
z^4(1 - 30 z^4 - 40z^6) \ln^2  \frac{y_1}{y_2} \no \\  
& \hspace{-3.3cm}    
- 576 z^4 F(z) +8z^2(2y_1 + 1)(6 + 11z^2 - 70z^4 -120 z^6) \ln
\frac{y_1}{y_2} \bigg\} \, a_2^{M_2},  
\label{eq:convs}
\end{align}
where we defined
\begin{align}
y_{1} \equiv \frac12 \left( \sqrt{1+4z^2} - 1 \right), \qquad 
y_{2} \equiv \frac12 \left( \sqrt{1+4z^2} + 1 \right), \hspace{2cm}\no
\\ 
F(z) \equiv \Lic(-y_1)-\Sab(-y_1)-\ln y_1 \Lib(-y_1)+\frac12 \ln
y_1\ln^2 y_2 -\frac{1}{12} \ln^3 z^2 + \zeta_3. 
\end{align}
In the massless limit the function $H_4(z_f)$ simplifies to
\begin{align}
H_4(0) &=\frac{22}{3}+7 a_1^{M_2} +\frac35 a_2^{M_2}.
\end{align}
The finiteness of the convolution integrals in (\ref{eq:convs})
completes the explicit factorization proof of the imaginary part of the
NNLO vertex corrections.    

We summarize our results for the vertex corrections in the Gegenbauer
representation of the light-cone distribution amplitude of the meson
$M_2$ (with $C_F=\frac43$, $C_A=3$, $n_f=5$)      
\begin{align}
\frac{1}{\pi} \; \Im \;\tilde{V}^{(1)}
    &=-3-3 a_1^{M_2},\no \\
\frac{1}{\pi} \; \Im \;\tilde{V}_1^{(2)}
    &=\left(9+9 a_1^{M_2}\right) \, \ln \frac{\mu^2}{m_b^2} +
    \frac{1189}{24} + \frac{47\pi^2}{90} -8 \zeta_3 -\left( \frac{33}{8}
      - \frac{23\pi^2}{10}  \right) a_1^{M_2} \no \\ 
&\quad - \left( \frac{173}{60} + \frac{9\pi^2}{35}\right) a_2^{M_2} ,\no\\ 
\frac{1}{\pi} \; \Im \;\tilde{V}_2^{(2)}
    &=-\left(26+ \frac{110}{3} a_1^{M_2}\right) \,  \ln
    \frac{\mu^2}{m_b^2} -\frac{10315}{72} +\frac{674\pi^2}{135}
    -\frac{28}{3} \zeta_3 \no \\ 
&\quad  -\left( \frac{10793}{72} - \frac{166\pi^2}{15} \right) a_1^{M_2}
-\left( \frac{3155}{48} -\frac{187\pi^2}{42} \right) a_2^{M_2} + H_4(z)
+ H_4(1), 
\label{eq:V12Gegen}
\end{align}
with $H_4(z_f)$ given in (\ref{eq:convs}). In order to illustrate the
relative importance of the individual contributions, we set $\mu=m_b$
and $z=m_c/m_b=0.3$ which yields  
\begin{align}
\Im \;\tilde{V}^{(1)}
    &= -9.425 - 9.43 a_1^{M_2},\no \\
\Im \;\tilde{V}_1^{(2)}
    &= 141.621 + 58.36  \; a_1^{M_2} - 17.03 \; a_2^{M_2} ,\no \\
\Im \;\tilde{V}_2^{(2)}
    &= -317.940 - 115.62 \; a_1^{M_2} - 68.31 \; a_2^{M_2}.
\label{eq:V12num}
\end{align}
We thus find large coefficients for the NNLO vertex corrections and
expect only a minor impact of the higher Gegenbauer moments, in
particular in the symmetric case with $a_1^{M_2}=0$. Notice that all
contributions add constructively in $\alpha_{1,2}$ due to the relative
signs of the Wilson coefficients, $\tilde{C}_1\sim1.1$ and
$\tilde{C}_2\sim-0.2$. In the case of $\alpha_1$ the contribution from
$\tilde{V}_1^{(2)}$ is found to exceed the formally leading contribution
$\tilde{V}^{(1)}$  due to the fact that the latter is multiplied by the
small Wilson coefficient $\tilde{C}_2$. Concerning $\alpha_2$ the NNLO
vertex corrections are also substantial, roughly saying they amount to a
$50\%$ correction. In both cases the impact of the charm quark mass is
small, we find a correction of $\sim3\%$ compared to the massless
case. A more detailed numerical analysis including the contributions
from spectator scattering will be given in the following section.   

We finally remark that the large $\beta_0$-limit considered
in~\cite{betazero} fails to reproduce the imaginary part of $\alpha_1$
as it completely misses the leading contribution from
$\tilde{V}_1^{(2)}$. In the case of $\alpha_2$ the approximation turns
out to be reasonably good with a deviation of $\sim10\%$ compared to the
full NNLO result.


\section{Numerical analysis}

\label{sec:numerics}

\subsection{Implementation of Spectator Scattering}
\label{sec:SpecScat}

In the numerical analysis we combine our results with the NNLO
corrections from 1-loop spectator scattering obtained
in~\cite{SpecScat:NLO:BJ05,SpecScat:NLO,SpecScat:NLO:BJ06,Jet,Jet:BY05}.
In contrast to the vertex corrections considered in this work, the
spectator term receives contributions from the hard scale $\mu_h\sim
m_b$ and the hard-collinear scale $\mu_{hc}\sim (\LQCD
m_b)^{1/2}$. According to this, the kernels $T_i^{II}$ from
(\ref{eq:QCDF}) factorize into hard functions $H_i^{II}$ and a (real)
hard-collinear jet-function $J_{||}$. Evaluating both kernels at the
same scale $\mu$ would imply parametrically large logarithms which may
spoil the convergence of the perturbative expansion.     

In order to resum these logarithms we follow
reference~\cite{SpecScat:NLO:BJ05} and perform the substitution  
\begin{align} 
& C_{i}(\mu) \;\,
  T_{i}^{II}(\mu) \otimes
  [\hat{f}_{B} \phi_B](\mu) \otimes
  \phi_{M_1}(\mu) \otimes
  \phi_{M_2}(\mu) \no \\
& \rightarrow \;
  C_{i}(\mu_h) \;\,
  H_{i}^{II}(\mu_h) \otimes
  {\cal{U}}_{||}(\mu_h,\mu_{hc}) \otimes
  J_{||}(\mu_{hc}) \otimes
  [\hat{f}_{B} \phi_B](\mu_{hc}) \otimes
  \phi_{M_1}(\mu_{hc}) \otimes
  \phi_{M_2}(\mu_h),
\label{eq:SpecEvol}
\end{align}
where ${\cal{U}}_{||}=e^{-S}\,U_{||}$ consists of a universal Sudakov
factor $S$ and a non-local evolution kernel $U_{||}$. As an imaginary 
part is first generated at $\calO(\as^2)$ in the spectator term, we
implement the resummation in the leading-logarithmic (LL)
approximation. In the traditional operator basis from
(\ref{eq:TradBasis}) the respective imaginary part takes the form       
\begin{align}
\Im \; \alpha_i(M_1 M_2) \big{|}_\text{spec}&= 
\frac{\as(\mu_h) \as(\mu_{hc}) C_F}{4N_c^2} \; 
\frac{9 f_{M_1} \hat{f}_B(\mu_{hc})}{m_b F_+^{B M_1}(0)
  \lambda_B(\mu_{hc})} \; \sum_{n,m} \, 
a_m^{M_1}(\mu_{hc}) \,a_n^{M_2}(\mu_h)\no \\
& \quad \times \Im  
\bigg[ \tilde{C}_{i\pm1}(\mu_h) \,\tilde{R}_1^{mn}(\mu_h,\mu_{hc}) +
\tilde{C}_i(\mu_h) \,\tilde{R}^{mn}_2 (\mu_h,\mu_{hc})
\bigg] + \calO(\as^3), 
\label{eq:alpha12Spec}
\end{align}
where we made the scale dependence of the parameters explicit and
introduced the first inverse moment of the $B$ meson light-cone
distribution amplitude $\lambda_B^{-1}$. We further wrote $1=a_0^M(\mu)$
in order to simplify the notation. In (\ref{eq:alpha12Spec}) the
resummation is encoded in  
\begin{align}
\tilde{R}_i^{mn}(\mu_h,\mu_{hc}) &=
\frac19 \,e^{-S(\mu_h,\mu_{hc})} 
\int_0^1 du \; 6 u \ubar \,C_n^{(3/2)}(2u-1) 
\int_0^1 dz \; \calC_m(z;\mu_h,\mu_{hc}) \,r_i(u,z),
\label{eq:Rinm}
\end{align}
with the Sudakov factor $S$ given in LL approximation in equation (106)
of~\cite{Jet:BY05} and the kernels $r_i$ in equations (38) and (39)
of~\cite{SpecScat:NLO:BJ05}. Following~\cite{SpecScat:NLO:BJ06} we
defined          
\begin{align}
\calC_m(z;\mu_h,\mu_{hc}) &=
\int_0^1 dv \; 6v \,C_m^{(3/2)}(2v-1) \, U_{||}(\bar v, \bar z; \mu_h,\mu_{hc}),
\end{align}
which can be computed by solving numerically the integro-differential
equation  
\begin{align}
\frac{d}{d\ln\mu} \calC_m(z;\mu,\mu_{hc}) &=
-\int_0^1 dw \; \gamma_{||}(\bar z, \bar w) \;
\calC_m(w;\mu,\mu_{hc})
\end{align}
with initial condition
$\calC_m(z;\mu_{hc},\mu_{hc})=6z\,C_m^{(3/2)}(2z-1)$ and $\gamma_{||}$
from equation (99) of~\cite{Jet:BY05}.

\begin{table}[b!]
\centerline{
\parbox{13cm}{\setlength{\doublerulesep}{0.1mm}
\centerline{\begin{tabular}{|l||c|c|c|c|c|c|c|c|c|} \hline
\hspace*{1cm}&\hspace*{1cm}&\hspace*{1cm}&\hspace*{1cm}&
\hspace*{1cm}&\hspace*{1cm}&\hspace*{1cm}&\hspace*{1cm}&
\hspace*{1cm}&\hspace*{1cm} 
\\[-0.7em]
& $\tilde{R}_i^{00}$ & $\tilde{R}_i^{01}$ & $\tilde{R}_i^{02}$
& $\tilde{R}_i^{10}$ & $\tilde{R}_i^{11}$ & $\tilde{R}_i^{12}$
& $\tilde{R}_i^{20}$ & $\tilde{R}_i^{21}$ & $\tilde{R}_i^{22}$ 
\\[0.3em]
\hline\hline\hline&&&&&&&&&
\\[-0.7em]
$i=1$ (I) & 
$11.0$ & $23.2$ & $29.4$ &
$14.1$ & $23.8$ & $30.8$ &
$15.1$ & $23.8$ & $31.3$ 
\\[0.3em]
\hline&&&&&&&&&
\\[-0.7em]
$i=1$ (II) & 
$9.88$ & $20.9$ & $26.5$ &
$12.5$ & $21.5$ & $27.8$ &
$13.4$ & $21.5$ & $28.2$ 
\\[0.3em]
\hline\hline\hline&&&&&&&&&
\\[-0.7em]
$i=2$ (I) & 
$-5.29$ & $-8.43$ & $-8.24$ &
$-6.58$ & $-11.0$ & $-11.3$ &
$-7.04$ & $-12.0$ & $-12.6$ 
\\[0.3em]
\hline&&&&&&&&&
\\[-0.7em]
$i=2$ (II) & 
$-4.77$ & $-7.60$ & $-7.43$ &
$-5.85$ & $-9.72$ & $-9.98$ &
$-6.27$ & $-10.6$ & $-11.2$ 
\\[0.3em]
\hline
\end{tabular}}
\vspace{4mm} \caption{\label{tab:Rimn}\small \textit{Numerical values of
  the imaginary part of $\tilde{R}_i^{mn}(\mu_h,\mu_{hc})$ from
  (\ref{eq:Rinm}) for $\mu_h=\mu_{hc}$ (line I, without resummation) and
  $\mu_h=4.8$GeV, $\mu_{hc}=1.5$GeV (line II, with resummation).}}}} 
\end{table}

In order to illustrate the numerical importance of the resummation we
compare the values of the imaginary part of $\tilde{R}_i^{mn}$ for
$m,n\leq2$ and $\mu_h=\mu_{hc}$ (line I, without resummation) and
$\mu_h=4.8$GeV, $\mu_{hc}=1.5$GeV (line II, with resummation) in
Table~\ref{tab:Rimn}. We observe that the resummation leads to a
suppression of the spectator term of $\sim10\%$ due to the universal
Sudakov factor ($e^{-S}\simeq0.89$ for our choice of input
parameters). The resummation effects induced by $U_{||}$ turn out to be
of minor numerical importance.        

According to (\ref{eq:alpha12Spec}) we must evolve the Gegenbauer
moments of the mesons $M_1$ and $M_2$ to the hard-collinear and the hard
scale, respectively. In LL approximation the Gegenbauer moments do not
mix and the evolution reads 
\begin{align}
a_i^{M}(\mu) = \left( \frac{\as(\mu_0)}{\as(\mu)}
\right)^{\gamma_i/2\beta_0}a_i^{M}(\mu_0)
\label{eq:EvolGegen}
\end{align}
with anomalous dimensions $\gamma_1=-\frac{64}{9}$ and
$\gamma_2=-\frac{100}{9}$.  

We are left with the evolution of the $B$ meson parameters to the
hard-collinear scale. We convert the HQET decay constant $\hat{f}_B$
into the physical one $f_B$ using the LL relation   
\begin{align}
\hat{f}_B(\mu) = \left( \frac{\as(\mu)}{\as(m_b)} \right)^{-2/\beta_0}
\, f_B.
\end{align}
The evolution of $\lambda_B$ is more complicated. The solution of the
integro-differential equation, which governs the LL evolution of the $B$
meson light-cone distribution amplitude, can be found
in~\cite{BmesonDA}. Here we adopt a model-description for the $B$ meson
distribution amplitude to generate the evolution of $\lambda_B$. We take
the model from~\cite{BmesonDA} which has the correct asymptotic
behaviour and is almost form-invariant under the evolution.     

Finally we implement the BBNS model from~\cite{BBNS} in order to
estimate the size of power corrections to the factorization
formula~(\ref{eq:QCDF}). This results in an additional contribution from
spectator scattering related to subleading projections on the
light-cone distribution amplitudes of the light mesons. It is given by  
\begin{align} 
\Im \; \alpha_i(M_1 M_2) \big{|}_\text{power}&= 
\frac{\pi \as C_F}{N_c^2} \; 
\frac{3 f_{M_1} \hat{f}_B}{m_b F_+^{B M_1}(0) \lambda_B} \;\;  
\tilde{C}_{i\pm1} \, r_\chi^{M_1} \, 
\Delta_{M_2} \;\Im [ X_H ],
\label{eq:alpha12Twist3}
\end{align}
where $r_\chi^{M}(\mu)=2m_{M}^2/\bar{m}_b(\mu)/(\bar{m}_q+\bar{m}_{\bar
  q})(\mu)$, $\Delta_{M}=1+\sum_{n} (-1)^n a_n^{M}$ and $X_H$
parameterizes an endpoint-divergent convolution integral. The latter is
written as      
\begin{align} 
X_H = (1 + \rho_H \,e^{i \varphi_H}) \, \ln \frac{m_B}{\Lambda_h}
\end{align}
which may generate an imaginary part due to soft rescattering of the
final state mesons. We take $\ln m_B/\Lambda_h\simeq2.3$,
$\rho_H=1$ and allow for an arbitrary phase $\varphi_H$.  

\subsection{Tree amplitudes in NNLO}

\begin{table}[t!]
\centerline{
\parbox{13cm}{\setlength{\doublerulesep}{0.1mm}
\centerline{\begin{tabular}{|c|c||c|c|} \hline
\hspace*{3.1cm}&\hspace*{3.1cm}&\hspace*{3.1cm}&\hspace*{3.1cm} \\[-0.7em]
Parameter & Value & Parameter & Value\\[0.3em]
\hline\hline\hline&&& 
\\[-0.7em]
$\Lambda_\text{\tiny\MSbar}^{(5)}$ & 
$0.225$ & 
$f_\pi$ & 
$0.131$
\\[0.3em]
$m_b$ & 
$4.8$ &
$f_B$ & 
$0.21\pm 0.02$
\\[0.3em]
$m_c$ & 
$1.6\pm 0.2$ & 
$F_+^{B\pi}(0)$ & 
$0.25 \pm 0.05$
\\[0.3em]
$\bar{m}_b(\bar{m}_b)$ & 
$4.2$ & 
$\lambda_B(1\gev)$ & 
$0.48 \pm 0.12$
\\[0.3em]
$(\bar{m}_u\!+\!\bar{m}_d)(2\gev)$ & 
$0.008\pm0.002$ & 
$a_2^\pi(1\gev)$ & 
$0.25\pm 0.2$
\\[0.3em]
\hline
\end{tabular}}
\vspace{4mm} \caption{\label{tab:numeric}\small \textit{Theoretical input
    parameters (in units of \gev~or dimensionless).}}}} 
\end{table}

The numerical implementation of the vertex corrections from
(\ref{eq:alpha12Trad}) is easier since they can be evaluated at the
hard scale $\mu_h\sim m_b$ and depend only on few parameters. As can be
read off from (\ref{eq:V12Gegen}), the first Gegenbauer moment of the
meson $M_2$ enters the leading term $\tilde{V}^{(1)}$. We therefore
require its next-to-leading logarithmic (NLL) evolution (which can be
found in~\cite{PionNLL}) but as we restrict our attention to
$B\to\pi\pi$ decays in the following the first moment does not
contribute at all. Since the second Gegenbauer moment does not enter
$\tilde{V}^{(1)}$, it is only required in LL approximation given by
(\ref{eq:EvolGegen}).           

Our input parameters for the $B\to\pi\pi$ tree amplitudes are summarized
in Table~\ref{tab:numeric}. The value for the $B$ meson decay constant
is supported by QCD sum rule calculations~\cite{fBSumRules} and recent
lattice results~\cite{fBLattice}. The form factor $F_+^{B\pi}$ (at large
recoil) has been addressed in the light-cone sum rule (LCSR)
approach~\cite{FplusLCSR}. As we implement the model
from~\cite{BmesonDA} for the $B$ meson distribution amplitude, we take
the respective value for $\lambda_B$. This value is somewhat larger than
the one used in previous QCD Factorization
analyses~\cite{BBNS,SpecScat:NLO:BJ05,SpecScat:NLO,SpecScat:NLO:BJ06},
but it is supported by a QCD sum rule and a LCSR
calculation~\cite{lambdaBSumRules}. The value for the second Gegenbauer
moment of the pion can be inferred from a LCSR analysis~\cite{a2LCSR}
and lattice results~\cite{a2lattice}.    

In order to estimate the size of higher-order perturbative corrections
we vary the hard scale in the range $\mu_h=4.8^{+4.8}_{-2.4}~\gev$ and
the hard-collinear scale independently between
$\mu_{hc}=1.5^{+0.9}_{-0.5}~\gev$. Throughout we use 2-loop running of
$\as$ with $n_f=5$ ($n_f=4$) for quantities that are evaluated at the
hard scale $\mu_h$ (hard-collinear scale $\mu_{hc}$). The quark masses
are interpreted as pole masses except for those entering $r_\chi^M$.

With these input parameters the complete NNLO result for the imaginary
part of the topological tree amplitudes is found to be
\begin{align}
\Im\; \alpha_1(\pi\pi)& \,=\,~~ 0.012 \big{|}_{V^{(1)}}
                            + 0.031 \big{|}_{V^{(2)}}
                            - 0.012 \big{|}_{S^{(2)}} \no \\
                      & \,=\,~~ 0.031 \pm 0.015~(\text{scale})
                                    \pm 0.006~(\text{param})
                                    \pm 0.010~(\text{power})\no\\
                      & \,=\,~~ 0.031 \pm 0.019,\no\\
\Im\; \alpha_2(\pi\pi)& \,=   - 0.077 \big{|}_{V^{(1)}}
                            - 0.052 \big{|}_{V^{(2)}}
                            + 0.020 \big{|}_{S^{(2)}} \no \\
                      & \,=   - 0.109 \pm 0.023~(\text{scale})
                                    \pm 0.010~(\text{param})
                                    \pm 0.045~(\text{power})\no \\
                      & \,=   - 0.109 \pm 0.052.
\label{eq:NNLOresult}
\end{align}
In these expressions we disentangled the contributions from the NLO
(1-loop) vertex corrections $V^{(1)}$, NNLO (2-loop) vertex corrections
$V^{(2)}$ and NNLO (1-loop) spectator scattering $S^{(2)}$. In both
cases the NNLO corrections are found to be important, although small in
absolute terms. For the imaginary part of $\alpha_1$ the NNLO
corrections exceed the formally leading NLO result which can be
explained by the fact that the latter is multiplied by the small Wilson
coefficient $\tilde{C}_2$, cf.~the discussion after
(\ref{eq:V12num}). We further observe that the individual NNLO
corrections come with opposite signs which leads to a partial
cancellation in their sum. The phenomenologically most important
consequence of our calculation may be the enhancement of the imaginary
part of the colour-suppressed tree amplitude $\alpha_2$.

In our error estimate in (\ref{eq:NNLOresult}) we distinguished between
uncertainties which originate from the variation of the hard and the
hard-collinear scale (scale), from the variation of the input parameters
in Table~\ref{tab:numeric} (param) and from the BBNS model which we used
to estimate the size of power-corrections (power). By now we expect the
power corrections to be the main limiting factor for an accurate
determination of the amplitudes. However, although the inclusion of NNLO
corrections has reduced the dependence on the renormalization scales, it
still remains sizeable (in particular for $\mu_h$). For our final error
estimate in the third line of each amplitude, we added all uncertainties
in quadrature.       

Finally we remark that our numerical values for the NNLO spectator terms
are much smaller than the ones quoted in~\cite{SpecScat:NLO:BJ05}. This
is partly related to the fact that we use different hadronic input
parameters, in particular a much larger value for $\lambda_B$. In
addition to this, the authors of~\cite{SpecScat:NLO:BJ05} essentially
evaluate the hard functions $H_i^{II}$ at the hard-collinear scale in
order to partly implement the (unknown) NLL resummation of
parametrically large logarithms. The NLL approximation is indeed
required for the real part of the amplitudes, but as long as we
concentrate on the imaginary part it is consistent to work in the LL
approximation as discussed in Section~\ref{sec:SpecScat}. As the
spectator term is the main source for the uncertainties from the
hadronic input parameters, we also obtain smaller error bars
than~\cite{SpecScat:NLO:BJ05}.

\newpage

\section{Conclusion}

\label{sec:conclusion}

We computed the imaginary part of the 2-loop vertex corrections to the
topological tree amplitudes in charmless hadronic $B$ decays. Together
with the 1-loop spectator scattering contributions considered
in~\cite{SpecScat:NLO:BJ05,SpecScat:NLO,SpecScat:NLO:BJ06,Jet,Jet:BY05},
the imaginary part of the tree amplitudes is now completely determined
at NNLO in QCD Factorization. 

Among the technical issues we showed that soft and collinear infrared
divergences cancel in the hard-scattering kernels and that the resulting
convolutions are finite, which demonstrates factorization at the 2-loop
order. In our numerical analysis we found that the NNLO corrections are
significant, in particular they enhance the strong phase of the
colour-suppressed tree amplitude $\alpha_2$. Further improvements of the
calculation still require a better understanding of power corrections to
the factorization formula.  

Our calculation represents an important step towards a NNLO prediction
of direct CP asymmetries in QCD Factorization. As the topological
penguin amplitudes, which also affect the direct CP asymmetries, have
not yet been computed completely at NNLO (the contribution from
spectator scattering can be found in~\cite{SpecScat:NLO:BJ06}), we
refrained from discussing them already in this work. Moreover, the
strategy outlined in this work may also be applied for the calculation
of the real part of the topological tree amplitudes, which is, however,
technically more involved~\cite{GB:thesis,GB:RePart}.

  
\subsection*{Acknowledgements}


It is a pleasure to thank Gerhard Buchalla for his continuous help and
guidance and for helpful comments on the manuscript. I am also grateful
to Volker Pilipp and Sebastian J\"ager for interesting discussions. This
work was supported in part by the German-Israeli Foundation for
Scientific Research and Development under Grant G-698-22.7/2001 and by
the DFG Sonderforschungsbereich/Transregio 9.


\end{document}